\documentclass[journal,twocolumn]{IEEEtran}

\usepackage{cite}
\usepackage{amsmath,amssymb,amsfonts}
\usepackage{multirow}
\usepackage{graphicx}
\usepackage{subfig}
\usepackage{textcomp}
\usepackage{tablefootnote}
\usepackage[utf8]{inputenc}
\usepackage[english]{babel}
\usepackage{fancyhdr}
\usepackage[font=small,skip=0pt]{caption}
\usepackage{array}
\usepackage{tabularx}
\usepackage[ruled,vlined]{algorithm2e}
\usepackage{algpseudocode}
\usepackage[section]{placeins}
\usepackage{float}
\usepackage{amsmath}
\usepackage{optidef}
\usepackage{amsthm}
\usepackage[dvipsnames]{xcolor}

\newtheorem{remark}{Remark}

\newtheorem{definition}{Definition}

\def\BibTeX{{\rm B\kern-.05em{\sc i\kern-.025em b}\kern-.08em
		T\kern-.1667em\lower.7ex\hbox{E}\kern-.125emX}}
\begin{document}
	
	\title{Power Control with QoS Guarantees: A Differentiable Projection-based  Unsupervised  Learning Framework \\
	}
	\author{Mehrazin Alizadeh and~Hina~Tabassum, {\em Senior Member IEEE}
		\thanks{ This research is supported by a Discovery Grant funded by the Natural Sciences and Engineering Research Council of Canada. M. Alizadeh and H.~Tabassum are with the Department of Electrical Engineering and Computer Science, York University, ON, Canada (E-mail: mehrazin@eecs.yorku.ca,  hinat@yorku.ca).}
		\vspace{-8mm}
	}
	
	\raggedbottom
	
	\maketitle
	
	\begin{abstract}
		\\
		Deep neural networks (DNNs) are   emerging as a potential solution to solve NP-hard wireless resource allocation problems.  However, in the presence of intricate constraints, e.g., users' quality-of-service (QoS) constraints, guaranteeing constraint satisfaction becomes a fundamental challenge. In this paper, we propose a novel unsupervised learning framework to solve the classical power control problem in a multi-user interference channel, where the objective is to maximize the network sum-rate \textcolor{black}{under users' minimum data rate or QoS requirements} and power budget constraints. 
		Utilizing a differentiable projection function,  two novel deep learning (DL) solutions are pursued. The first is called Deep Implicit Projection Network (DIPNet), and the second is called Deep Explicit Projection Network (DEPNet). DIPNet utilizes a differentiable convex optimization layer to implicitly define a projection function. On the other hand, DEPNet uses an explicitly-defined projection function, which has an iterative nature and relies on a differentiable correction process. DIPNet requires convex constraints; whereas, the DEPNet does not require convexity and has a reduced computational complexity.
		To enhance the sum-rate performance of the proposed models even further, Frank-Wolfe algorithm (FW)  has been applied to the output of the proposed models.
		Extensive simulations depict that the proposed DNN solutions not only improve the achievable data rate but also achieve zero constraint violation probability, compared to the existing DNNs. The proposed solutions outperform the classic optimization methods  in terms of computation time complexity.

	\end{abstract}
	\begin{IEEEkeywords}
		Power control, learning to optimize (L2O), deep learning (DL), unsupervised learning, differentiable projection, multi-user, interference, and resource allocation.
	\end{IEEEkeywords}
	
	\section{Introduction}
	
	The problem of sum-rate maximization (SRM) in a multi-user interference channel through optimized power control has been explored for decades using standard optimization tools. However, due to the non-convex and NP-hard  nature of the power control problem and lack of analytical solutions, a majority of the existing  algorithms rely on an either exhaustive search
	(explicitly or implicitly) \cite{liu2012achieving} or iterative optimization of some approximate sub-problems \cite{shi2011iteratively}. The convergence and computational complexity typically hinder the practicality of the optimal or near-optimal solutions \cite{liang2019towards}. One way to mitigate the computational complexity of solving NP-hard optimization problems  is to view  them as a mapping from the state of the environment to the decision variables. This mapping can be learned efficiently by deep neural networks (DNNs) via offline training. Since the inference time of DNNs is far less than the run-time of iterative algorithms, online computational complexity will reduce significantly. 
	
	While DNNs can  minimize the time complexity, handling sophisticated problem constraints is a fundamental challenge regardless of the supervised or unsupervised training method. Note that, simple constraints (e.g., power budget constraint in a power control problem \cite{liang2019towards}, base station (BS) quota constraint in a user assignment problem \cite{kaushik2021deep}, etc.) can be satisfied using standard activation functions (like Rectified Linear Unit (ReLU), Sigmoid, etc.). Nevertheless, sophisticated  quality-of-service (QoS){\footnote{ \textcolor{black} {In this paper, the term QoS refers to  users' minimum data rate requirements which can be different due to distinct services required by those users.}}} constraints cannot be incorporated  using well-known activation functions. To date, such constraints are either 
	incorporated by considering a  penalty of the constraint violation into the loss function, which encourages the DNN output to meet the constraints \cite{liang2019towards, kaushik2021deep,added1} or included in the loss function which is the Lagrangian of the original problem and the learnable dual variables penalize the violation of constraints \cite{sun2019learning, added}. The approaches however do not provide a guarantee that the results are always feasible and satisfy constraints. Given the infancy of this line of research, this paper aims to address the following fundamental questions: (\textit{1) How to systematically incorporate convex and/or non-convex constraints into the DNN architecture instead of incorporating them in the loss function? (2) How to ensure a zero constraint violation probability?}

	\begin{table*}[t]
		\begin{center}
			\caption{A comparative analysis of the existing machine learning frameworks for power control.}
			\label{table:more}\textcolor{black}{
				\begin{tabular}{|p{20mm}|p{20mm}|c|c|c|}
					\hline
					{\textbf{Ref.}} & {\textbf{QoS Violation}}& {\textbf{Constraint type}} & {\textbf{Constraints}}               & {\textbf{Method}}                              \\ \hline
					\cite{sun2018learning, deng2019application}                      & N/A & Linear & Power budget  & ReLU activation   \\ \hline 
					\cite{lee2018deep,shen2020graph}                      & N/A & Linear & Power budget  & Sigmoid activation    \\ \hline
					\cite{liang2019towards, added1}                      & Yes & Non-convex &  Power budget, QoS  & Customized loss function  \\ \hline  
					\cite{eisen2020optimal, sun2019learning, added}                    & Yes & Non-convex & Power budget, QoS  & Primal-Dual training   \\ \hline  
					\cite{naderializadeh2020wireless}                      & Yes & Non-convex & Power budget, QoS  & Counterfactual primal-dual learning    \\ \hline  
					\cite{li2021multicell}           & No & Linear & Power budget, QoS  & Heuristic closed-form projection  \\ \hline  
					\cite{9281322}                     & No$^{1}$  & Non-convex & Power budget, QoS & Customized loss function  \\ \hline  
					This paper                    & No & Non-convex & Power budget, QoS  & Differentiable implicit and explicit projection \\ \hline  
			\end{tabular}}              
		\end{center}
		\footnotesize{$^1$ \textcolor{black}{A heuristic ERP method has been applied to allocate powers and mitigate the QoS violation.}}
	\end{table*}

	\subsection{Background Work}

	
	
	To date, most deep learning (DL)-based studies considered SRM via power control \cite{sun2018learning, deng2019application, shi2011iteratively, shen2020graph}, while considering simple BS power budget constraint. 
	\textcolor{black}{Recently, some research works considered the problem of SRM via power control with power budget  and minimum rate constraint of users.  The authors in  \cite{liang2019towards, added1} applied unsupervised training of DNNs and incorporated the penalty of violating the QoS constraint in the loss function.  In \cite{9281322}, the authors proposed a hybrid resource allocation scheme for multi-channel underlay device-to-device (D2D) communications. In particular, the transmit power control is considered for SRM considering interference and minimum rate constraints. The authors considered a heuristic equally reduced power (ERP) scheme together with a DNN-based scheme to avoid violation of QoS constraints. However, resorting to these heuristics (ERP in \cite{9281322} or minimum power \cite{liang2019towards}), when the DNN's output violates QoS constraint,  can compromise the quality of optimal power allocation solutions; thereby impacting the  maximum achievable sum-rate.} 
	
	
	\textcolor{black}{Based on the duality theory, \cite{sun2019learning, added, eisen2020optimal} applied the Lagrangian loss function to train the DNN and parameterize both the primal and dual variables. However, due to the residual error of DNN for parameterizing the dual variables, the methods cannot guarantee that the constraints are always satisfied.}  In \cite{naderializadeh2020wireless}, the authors  introduced a slack variable that relaxes the minimum rate constraints. The objective changed to maximizing the sum-rate while minimizing the slack variable. This new method called counterfactual
	optimization sacrifices the sum-rate to provide QoS of the cell-edge users.   
	Very recently, \cite{li2021multicell}  studied the SRM via power control with minimum rate and power budget constraints for an ad-hoc setup. To address the constraints,  an approximate closed-form projection is used that takes the output of the DNN and projects it into the feasible set. However, the approach is limited to linear constraints. \textcolor{black}{\textbf{Table~I} summarizes the  existing state-of-the-art and clarifies the novelty of this article.}

	\subsection{Motivation and Contributions}
	Most of the aforementioned works handled the power budget constraint systematically via using a sigmoid (in case of a single channel) or softmax (in case of multiple channels) at the output layer of the DNN \cite{shen2020graph}. The QoS constraints, on the other hand, were handled either by adding a penalty term to the loss function, indicating the violation of these constraints, or by transforming the problem into its Lagrangian and performing the learning in the dual domain \cite{she2021tutorial, eisen2019dual}. Neither of those guarantees zero violation of constraints at the test time; therefore heuristic algorithms are generally applied to allocate feasible powers in the instance of QoS violation.
	In this paper, our contributions can be summarized as follows:
	\begin{itemize}
		\item We propose two novel DL solutions for the classical power control problem in a multi-user interference channel with QoS and power budget constraints. The first is called \textbf{D}eep \textbf{I}mplicit \textbf{P}rojection \textbf{Net}work (DIPNet), and the second is called \textbf{D}eep \textbf{E}xplicit \textbf{P}rojection \textbf{Net}work (DEPNet). The former requires convex constraints; whereas, the latter does not.

		\item DIPNet utilizes differentiable convex optimization layer \cite{amos2017optnet, agrawal2019differentiable}, a type of implicit layers in DNNs \cite{toturial_imp}, to implicitly define a projection function. The projection function projects the neural network's output to the feasible set defined by the QoS and power budget constraints. Thus, the DNN's output always satisfies the constraints. 
		

		\item DEPNet  is  inspired by \cite{donti2021dc3}, where a process called correction is applied iteratively on the output of the DNN to make it fall unto the feasible set of inequality constraints. The iterative process can be perceived as a differentiable and explicitly-defined projection function that projects the output of DNN to the feasible set of the problem with reduced computational complexity. The process is  compatible with GPU-based training.
		
		\item To improve the sum-rate performance of the proposed models even further, the Frank-Wolfe algorithm (FW), an iterative algorithm for constrained optimization problems \cite{lacoste2016convergence}, has been applied. This algorithm takes the output of the proposed models as its initial point and searches within the feasible set to find a better solution.
		
		\item The proposed models are trained in an unsupervised manner and compared with (i) the enhanced version of PCNet  \cite{liang2019towards}, called PNet, which works in the multi-channel scenario, is considered as the DNN-based benchmark, (ii) Geometric Program (GP) \cite{chiang2008power} and genetic algorithm are used as the optimization-based benchmark. The network sum-rate, constraint violation probability, and online test time are  the performance  metrics.
		
		\item Numerical results demonstrate that the proposed DIPNet and DEPNet guarantee zero constraint violation probability while outperforming PNet in terms of network sum-rate and constraint violation probability. In addition, the proposed models outperform  GP and genetic algorithm in terms of computation time complexity.
	\end{itemize}
	\textcolor{black}{Note that \cite{toturial_imp} and \cite{donti2021dc3} have proposed the fundamental techniques, i.e., deep implicit layers  and iterative gradient-descent-based projection, respectively. In this paper,  we have demonstrated  their applicability to handle both the convex and non-convex constraints in wireless resource allocation problems.}
		The proposed differentiable projection functions are compatible with any DNN architecture. 
		Thus, considering sophisticated  DNN architectures will not impact the functionality and applicability of the proposed projection methods.
		
		\subsection{Paper Organization and Notations}
		The remainder of this paper is organized as follows. Section~II details the system model, assumptions, and problem statement. Section~III depicts the problem transformation and introduces the proposed differentiable projection framework. Section~IV and Section~V detail the implicit and explicit projection framework. Section~VI proposes the Frank-Wolfe enhancement to the proposed DNN architectures. Section~VII details the experimental set-up, dataset generation procedure, and considered benchmarks for performance comparison. Section~VIII presents selected numerical results followed by conclusion in Section~IX.
		
		Throughout the paper, we use the following notations: bold lower-case letters for vectors, bold upper-case letters for matrix, and calligraphy upper-case letters denote a set. $\mathbb{R}$ and $\mathbb{C}$ denote the set of real and complex numbers. 
		
		\section{System Model and Problem Statement}
		\begin{figure}[t]
			\centering
			\includegraphics[scale=0.2,  trim=4 4 4 4 ,clip]{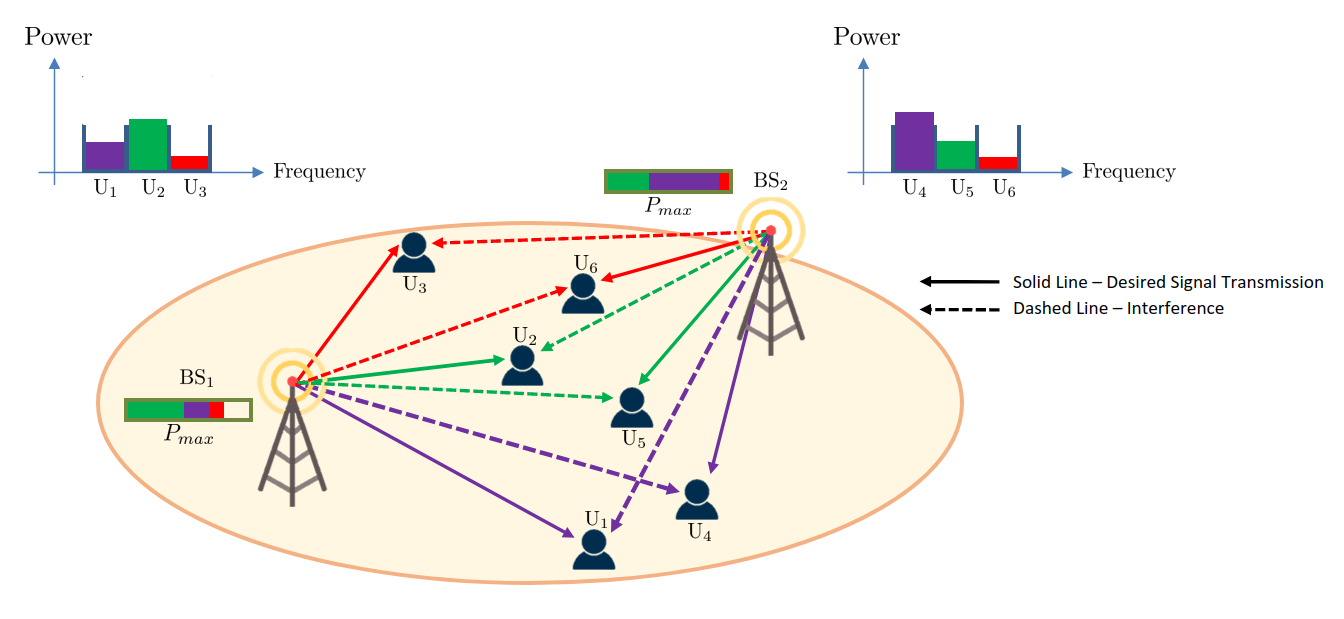}
			\caption{\textcolor{black}{A graphical illustration of the considered system model.}}
			\label{fig: NN-arc}
		\end{figure}
		We consider a downlink wireless network composed of $B$ single-antenna BSs where each BS can serve $Q$ users at maximum in $Q$ orthogonal frequency channels.  Due to orthogonal channel allocation at each BS, the users that are being served by the same BS will not interfere with each other, i.e., no intra-cell interference exists. However, the BSs share the same frequency spectrum and they equally distribute the bandwidth among their users, denoted by $W$. Thus, the inter-cell interference on each channel of bandwidth $W$ exists from the neighboring BSs. Without the loss of generality, we consider a total of $U$ single-antenna users in the system, where $U=BQ$. The achievable rate of the user associated with channel $q$ of BS $b$ can thus be modeled as follows:
		\begin{align}
			&R_{b,q}(\textbf{P}, \textbf{H}) = W \mathrm{log}_2 \left(1 + \gamma_{b,q}(\textbf{P}, \textbf{H})\right), \quad   \nonumber\\& \gamma_{b,q}(\textbf{P}, \textbf{H}) = \frac{H_{b,q,b} P_{b,q}}{\sum_{\hat{b}=1, \hat{b}\neq b}^{B} H_{b,q, \hat{b}} P_{\hat{b}, q} + \sigma^2}
		\end{align}
		where $H_{b,q,\hat{b}}$ denotes the interfering  channel between the  BS $\hat{b}$ and the user who is assigned to channel $q$ of BS $b$,  $P_{b,q}$ denotes the transmit power allocated   to the user scheduled on channel $q$ of BS $b$, and $\gamma_{b,q}$  denotes the received Signal-to-Interference-to-Noise ratio (SINR)  of the user scheduled on channel $q$ of BS $b$.
		Note that $\textbf{P} \in \mathbb{R}^{B \times Q}$ and $\textbf{H} \in \mathbb{R}^{B\times Q \times B}$ denote the matrix and tensor containing all values of the transmit powers and  channel power gains composed of distance-based path-loss, shadowing, and fading, respectively. We assume that the perfect channel state information (CSI) is available on the BS side. Also, $\sigma^2$ refers to the thermal noise power at the users' receivers, which is the same for all the users. \textcolor{black}{We denote the set of all BSs and channels as $\mathcal{B} = \{1,\cdots,B\}$ and $\mathcal{Q} = \{1,\cdots,Q\}$, respectively.}
		
		The SRM problem with QoS constraints  can then be formulated  as follows:
		\begin{equation}
			\begin{aligned} 
				\underset{\textbf{P}}{\text{maximize}}
				& \quad R(\textbf{P},\textbf{H}) =  \sum_{b = 1}^{B} \sum_{q = 1}^{Q} R_{b,q}(\textbf{P},\textbf{H}) \\
				\text{subject to}
				& \quad P_{b,q}\geq 0, \quad \forall b \in \mathcal{B}, \forall q \in \mathcal{Q}  \\
				& \quad \sum_{q = 1}^{Q} P_{b,q} \leq P_{\mathrm{max}}, \quad \forall b \in \mathcal{B} \\
				& \quad R_{b,q}(\textbf{P}, \textbf{H}) \geq \alpha_{b,q}, \quad \forall b \in \mathcal{B}, \forall q \in \mathcal{Q}\\
			\end{aligned}
			\label{prob:main}
		\end{equation}
		where the first constraint ensures non-negative power allocations and the second constraint refers to the  transmit power budget  of each BS. We assume the same maximum power budget $P_{\mathrm{max}}$ of each BS. The third constraint refers to the minimum rate requirement of the user scheduled on the channel $q$ of BS $b$ ($\alpha_{b,q}$). The problem in \eqref{prob:main} is  NP-hard and non-convex both in its objective and the constraints set; thus, finding an optimal solution is  challenging.

		\section{Problem Transformation and Differentiable Projection Framework}
		Since the implicit projection method, introduced in Section~IV, requires the convexity of the constraints and the explicit projection, introduced in Section~V, provides improved results when the feasible set is convex, we first transform the problem to its equivalent form with convex constraints. However, the non-convexity still arises from the non-convex objective function. 
		
		\subsection{Problem Transformation}
		The considered power control problem in \eqref{prob:main} can be reformulated in two different ways, i.e., either using the matrix version of power or using the  vector form of the power allocations. The matrix form of \eqref{prob:main} is shown below:
		\begin{equation}
			\begin{aligned} 
				\underset{\textbf{P}}{\text{maximize}}
				& \quad R(\textbf{P},\textbf{H}) \\
				\text{subject to}
				&  \quad \textbf{P} \geq \textbf{0},  \quad \textbf{P}.\textbf{1} \leq P_{\mathrm{max}}\textbf{1}, \quad \gamma_{b,q}(\textbf{P}, \textbf{H}) \geq \beta_{b,q} \\
			\end{aligned}
			\label{prob:main:mat}
		\end{equation}
		where $\textbf{1} = [1,1,\cdots,1]^T$ is a $B$-dimensional vector and $\beta_{b,q} = 2^{ \frac{\alpha_{b,q}}{W}} - 1$ is the minimum SINR to get the minimum rate requirement. To reformulate the problem in the vector form, we convert $\textbf{P}$ in a vector form. 
		The vector form, i.e. $\textbf{p} \in \mathbb{R}^{BQ \times 1}$, can be derived by stacking the $Q$ columns of matrix $\textbf{P}$, i.e., $P_{b,q} = p_{(q-1)B + b}$.
	The vector form of \eqref{prob:main:mat} is shown below:
	\begin{equation}
		\begin{aligned} 
			\underset{\textbf{p}}{\text{maximize}}
			& \quad R(\textbf{p},\textbf{H}) \\
			\text{subject to}
			& \quad \textbf{p} \geq \textbf{0}, \quad \textbf{A}\textbf{p} \leq P_{\mathrm{max}}\textbf{1}, \quad \textbf{C}\textbf{p} \geq \textbf{d}\\
		\end{aligned}
		\label{prob:main:vec}
	\end{equation}
	where $\textbf{A} \in \mathbb{R}^{B \times BQ}$ is defined as follows:
	\begin{equation}
		\centering
		\begin{aligned}
			A_{i,j} = 
			\left\{
			\begin{array}{ll}
				1  & \mbox{if } i \equiv j \quad (\text{mod } B) \\
				0 & \mbox{otherwise}
			\end{array} \right.
		\end{aligned}    
	\end{equation}
	The problem in \eqref{prob:main:vec} offers a linear (thus convex) formulation of the third constraint. 
	Since each BS can serve $Q$ users at a time and has a certain predefined users' quota,  the matrix $\textbf{C} \in R^{U \times U}$ becomes a block-diagonal matrix, where each block is related to one of the $Q$ channels. The $q$-th block denoted by $\textbf{M}^q \in \mathbb{R}^{B \times B}$ is defined as: 
	\begin{equation}
		\centering
		\begin{aligned}
			M^q_{b,\hat{b}} = 
			\left\{
			\begin{array}{ll}
				H_{b,q,b}  & \mbox{if } b = \hat{b} \\
				-\beta_{b, q}H_{b,q,\hat{b}} & \mbox{if } b \neq  \hat{b}
			\end{array} \right., 
		\end{aligned}    
	\end{equation}
	where $ \textbf{C} = \mathrm{diag}(\textbf{M}^1, ..., \textbf{M}^Q),$
	and $\textbf{d} \in \mathbb{R}^{U\times 1}$ is also derived by stacking the columns of matrix $\textbf{D} \in \mathbb{R}^{B \times Q}$, defined below, on top of each other, i.e.,
	$
	D_{b,q} = \beta_{b,q}\sigma^2, \quad \textbf{d} = [\textbf{D}_{:,1}^T, ..., \textbf{D}_{:,Q}^T]^T.
	$
	The aforementioned transformations also show that the third constraint can be expressed as either a non-convex, non-linear, or linear constraint in \eqref{prob:main}, \eqref{prob:main:mat}, and \eqref{prob:main:vec}, respectively \cite{chiang2008power}. In the following, we will leverage the aforementioned transformations in Section~IV and Section~V.
	\begin{figure}
		\centering
		\includegraphics[scale=0.25]{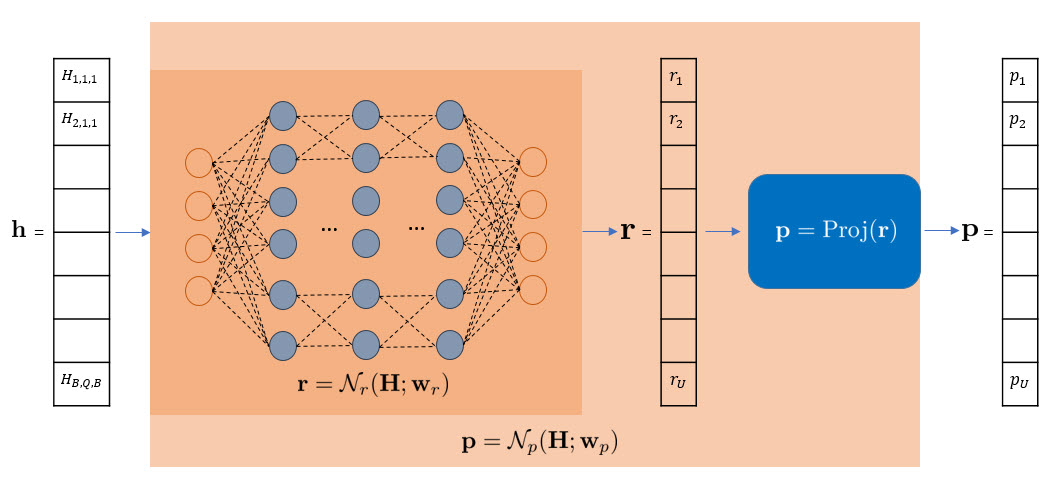}
		\caption{A graphical illustration of the proposed differentiable projection framework.}
		\label{fig: NN-arc}
	\end{figure}
	\subsection{Functional Optimization Form}
	Despite the  linearity of the constraints in \eqref{prob:main:vec}, this problem is  NP-hard due to the non-convex objective function. Thus, finding an optimal solution is  challenging. Traditionally,  \eqref{prob:main:vec} is solved  for each channel realization, i.e.,  for each realization of $\textbf{H}$, we solve \eqref{prob:main:vec} to get $\textbf{p}$ which dictates an implicit mapping between $\textbf{H}$ and $\textbf{p}$. Although  effective, this variable optimization approach yields high computational complexity. To overcome this problem, we can approximate the implicit mapping between $\textbf{H}$ and $\textbf{p}$  with an explicit function. This will significantly improve the computational complexity as long as the explicit function has an efficient implementation, e.g., using neural networks. 
	The equivalent functional optimization form of \eqref{prob:main:vec} is:
	\begin{equation}
		\begin{aligned}
			& \underset{F(\textbf{H})}{\text{maximize}}
			& & \mathbb{E}_{\textbf{H} \sim p(\textbf{H})}[R(F(\textbf{H}), \textbf{H})] \\
			& \text{subject to}
			& & F(\textbf{H}) \geq \textbf{0}, \quad \forall \textbf{H} \in \textcolor{black}{\mathbb{R}^{B \times Q \times B}}\\
			&&& \textbf{A}F(\textbf{H}) \leq P_{\mathrm{max}}\textbf{1}, \quad \forall \textbf{H} \in \textcolor{black}{\mathbb{R}^{B \times Q \times B}}\\
			&&& \textbf{C}F(\textbf{H})\geq \textbf{d}, \quad \forall \textbf{H} \in \textcolor{black}{\mathbb{R}^{B \times Q \times B}}
		\end{aligned}
		\label{prob:F-main-vec}
	\end{equation}
	where $F(\cdot)$ represents the functionality that maps CSI to a power allocation. It has been proven in \cite{sun2020unsupervised} that the solution of \eqref{prob:F-main-vec} is also the optimal solution of \eqref{prob:main:vec}. The same transition can be written for \eqref{prob:main:mat}. The output of $F(\cdot)$ is a matrix for \eqref{prob:main:mat} and a vector for \eqref{prob:main:vec}. 
	
	

	DNNs  have been shown to be a very rich family of parametric functions, in a sense that even a DNN with fully-connected layers (FCNN) has universal function approximation property \cite{LESHNO1993861}, and  has shown success in approximating the aforementioned mapping in supervised and unsupervised ways. Thus, we consider them  for approximating $F$, i.e. $F(\textbf{H}) = \mathcal{N}_{p}(\textbf{H}; \textbf{w}_p)$ where $\mathcal{N}_p$ is a DNN, and \textcolor{black}{$\textbf{w}_p \in \mathbb{R}^{D_{w_p}}$ is a  $D_{w_p}$ dimensional vector containing all trainable parameters, i.e., weights and biases, of the DNN.} The output of the DNN appears as a subscript to the DNN and its parameters. For example, if the output of the DNN is variable $\textbf{y}$, the DNN and its parameters are denoted by $\mathcal{N}_y$ and $\textbf{w}_y$, respectively. As a result, the problem of finding $\textbf{w}_p$ via learning can thus be formulated as:
	\begin{equation}
		\begin{aligned}
			& \underset{\textbf{w}_p}{\text{minimize}}
			& & \mathbb{E}_{\textbf{H} \sim p(\textbf{H})}[l(\mathcal{N}_p(\textbf{H}; \textbf{w}_p), \textbf{H})] \\
			& \text{subject to}
			&& \textbf{w}_p \in \textcolor{black}{\mathbb{R}^{D_{w_p}}}
		\end{aligned}
		\label{prob:learning}
	\end{equation}
	where $l$ is the loss function and measures how good the output of the neural network is for a given data $\textbf{H}$. Importantly, the design of the loss function is critical to solving the constrained optimization problem. Most of the current research typically incorporates the power budget constraint into the output of the DNN by using bounded activation functions like Sigmoid \cite{liang2019towards}. Other constraints are generally incorporated by customizing the loss function using the dual problem formulation. The downside of this choice is that there is no easy way to make sure the output of the neural network always meets the constraints and lies in the feasible set of problem \eqref{prob:main:vec}. To overcome this issue, in what follows, we present a differentiable projection-based framework that projects the DNN's output into the feasible set of \eqref{prob:F-main-vec}.
	
	\begin{figure}[!ht]
		\centering
		\begin{minipage}{0.25\textwidth}
			\includegraphics[scale=0.135]{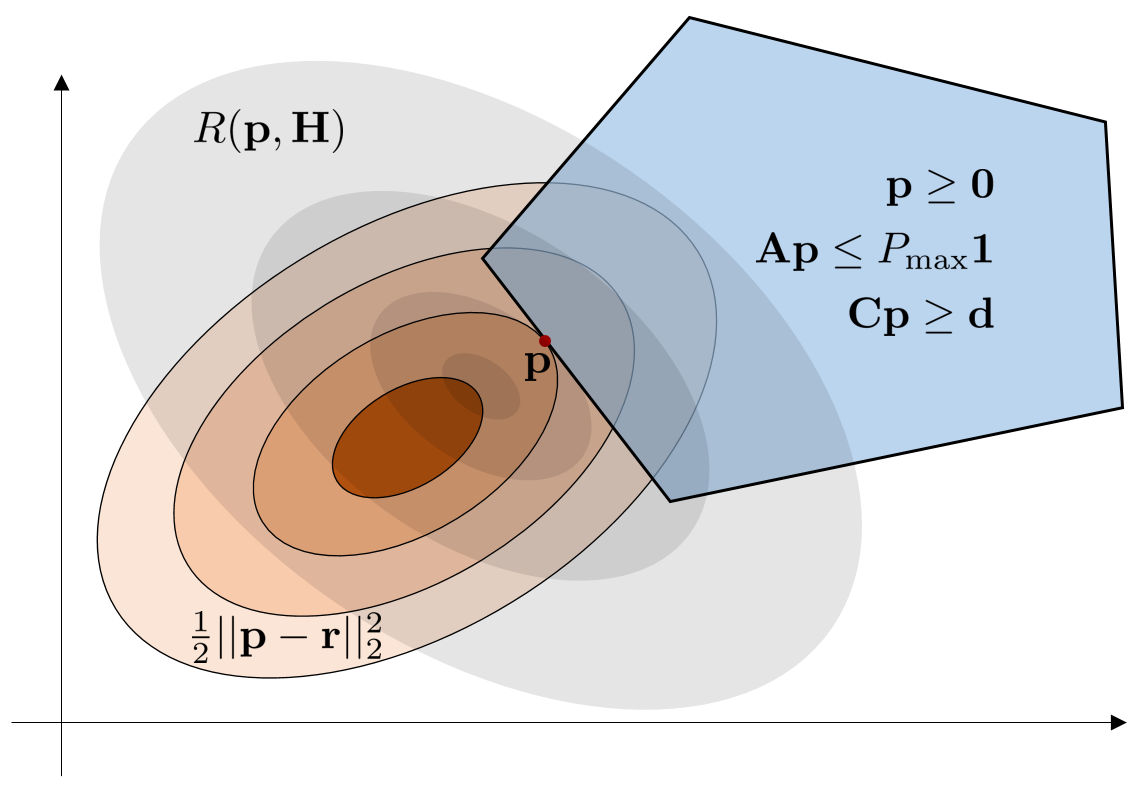}
		\end{minipage}\hspace{-0.3mm}
		\begin{minipage}{0.2\textwidth}
			\includegraphics[scale=0.135]{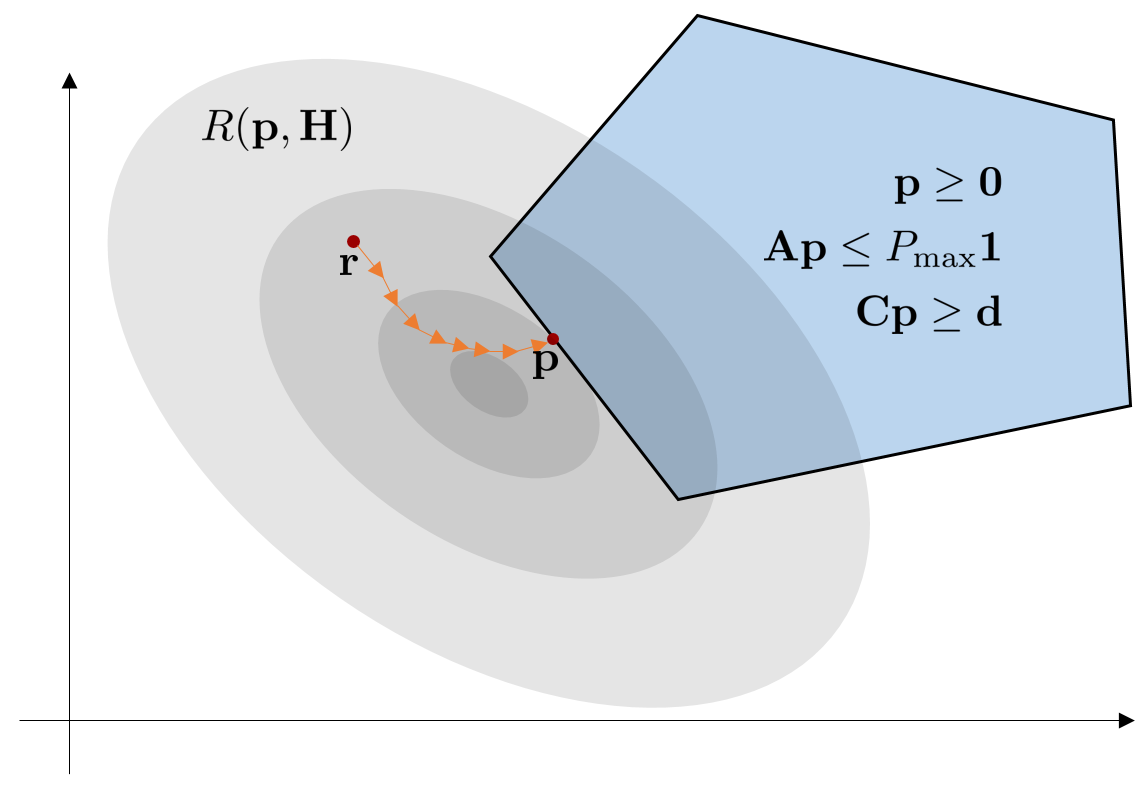}
		\end{minipage}
		\caption{An illustration of the proposed projection methods: Implicit projection via mathematical optimization (left) - Explicit projection via an iterative process (right).}
		\label{fig:projection}
	\end{figure}

	\subsection{Differentiable Projection Framework}
	In this paper, we  focus on designing  a projection framework where we can add a special layer to the DNN, which projects the output of the DNN to the feasible set of \eqref{prob:main:vec} or \eqref{prob:F-main-vec}. We refer to this transformation as \textit{Projection}, i.e., 
	\begin{equation}
		\textbf{p} = \mathcal{N}_p(\textbf{H};\textbf{w}_p) = \mathrm{Proj}(\textbf{r}); \quad \textbf{r} = \mathcal{N}_r(\textbf{H}; \textbf{w}_r),
		\label{diff-proj}
	\end{equation}
	where $\textbf{r}$ is the output of the backbone DNN without the projection layer, and $\mathrm{Proj}: \mathbb{R}^U \longrightarrow \mathbb{R}^U$ is a projection function unto the feasible set of \eqref{prob:F-main-vec}. Since we do not consider parametric projection functions, the parameters of $\mathcal{N}_p$ and $\mathcal{N}_r$ are considered as the same, i.e., $\textbf{w}_p = \textbf{w}_r$. Fig. \ref{fig: NN-arc} shows the graphical illustration of this architecture and the projection function is defined as follows. 
	\begin{definition}
		A function $\mathrm{Proj}: \mathbb{R}^U \longrightarrow \mathbb{R}^U$ is a differentiable projection function w.r.t. \eqref{prob:F-main-vec}, if its  Jacobian  can be evaluated, i.e., $\frac{\partial \textbf{p}}{\partial \textbf{r}}$ and $\textbf{p}$ meets the constraints of \eqref{prob:F-main-vec}. Being differentiable is critical to train the DNN end-to-end  using gradient-based methods\cite{lecun2015deep}.
	\end{definition}
	The projection function can be defined implicitly or explicitly as  in the following, respectively.

	\begin{itemize}
		\item \textbf{Differentiable Implicit Projection:} 
		in which a differentiable convex optimization (DCO) layer, a type of implicit layer in DNN, is applied to project the output of the $\mathcal{N}_r$ to the feasible set. By doing this, we make sure that the output of $\mathcal{N}_p$ always satisfies the constraints. The detailed explanation of DNN architecture is in Section~IV. Now that the output of DNN always satisfies the constraints, $l(\mathcal{N}_p(\textbf{H}; \textbf{w}_p), \textbf{H})$ can take the form of $-R(\mathcal{N}_p(\textbf{H}; \textbf{w}_p), \textbf{H})$ to directly optimize \eqref{prob:F-main-vec}.
		\item \textbf{Differentiable Explicit  Projection:}  uses a differentiable iterative process to realize the projection function (Proj) and moves the output of the $\mathcal{N}_r$ closer to the feasible set. Each iteration uses a process, called correction process \cite{donti2021dc3}, which \textit{corrects} the previous output towards lesser violation of the constraints. This approach uses soft-loss during training and shows faster performance relative to the first approach at the expense of the lack of the provable feasibility of the results. Experimental evaluations, however, confirm the zero constraint violation probability, as detailed in Section~VI.
	\end{itemize}
	\begin{remark}
		Let  $f: \mathbb{R}^m \longrightarrow \mathbb{R}^n$ be a function such that $\textbf{y} = f(\textbf{x})$. If the process of evaluating the output $\textbf{y}$ from an input $\textbf{x}$ is known, we refer to the function as \textbf{explicit}, whereas the \textbf{implicit} function means that the output evaluation process from the input is unknown. To put it differently, implicit definition separates "what" to compute from "how" to compute it \cite{gould2019deep}.
	\end{remark}
	It is noteworthy that the differentiable projection layer needs to consider both the power budget constraint and QoS constraints together, i.e., satisfy all constraints jointly.


\section{Differentiable Implicit Projection Framework}
In this section, we describe the systematic incorporation of the QoS constraint into the DNN architecture. To be more specific, we utilize the newly introduced  DCO layer \cite{agrawal2019differentiable} to project the output of the neural network to the feasible set defined by the QoS or minimum rate constraints. We call this layer a projection layer. DCO and other types of layers which describe an implicit functionality between input and output space lie under the umbrella of implicit layers \cite{toturial_imp}.

\subsection{Projection Layer}
We define the projection function implicitly using the concept of DCO which requires reformulating the constraints of the original problem as convex constraints and choosing a convex objective function for this layer. Using the affine formulation of the constraints of \eqref{prob:main}, as defined in \eqref{prob:main:vec} or \eqref{prob:F-main-vec}, the optimization problem that characterizes the projection layer is formulated as:
\begin{equation}
	\begin{aligned}
		& \textbf{p} =  \underset{\hat{\textbf{p}}}{\text{argmin}}
		\quad \frac{1}{2}||\hat{\textbf{p}} -  \textbf{r}||_2^2\\
		&\text{subject to}
		\qquad \hat{\textbf{p}} \geq \textbf{0}, \qquad  \textbf{A}\hat{\textbf{p}} \leq P_{\mathrm{max}}\textbf{1}, \qquad \textbf{C}\hat{\textbf{p}} \geq \textbf{d}\\
	\end{aligned}
	\label{prob:project-layer}
\end{equation}
where \eqref{prob:project-layer} implicitly defines the projection function ($\textbf{p} = \mathrm{Proj}(\textbf{r})$). This projection takes the form of the euclidean projection unto a set, defined by the constraints of \eqref{prob:F-main-vec}. Given the implicit function theorem \cite{toturial_imp, agrawal2019differentiable, gould2019deep}, the Jacobian of the output w.r.t. the input, i.e., $\frac{\partial \textbf{p}}{\partial \textbf{r}}$, can be computed, regardless of how the solution is derived. Moreover, since the constraints of \eqref{prob:project-layer} and \eqref{prob:F-main-vec} are the same, they have the same feasible set, i.e., the solution of \eqref{prob:project-layer}  satisfies the constraints of \eqref{prob:F-main-vec}.
Note that, as long as the convexity is preserved, one can choose other objective functions for \eqref{prob:project-layer} as well. The main role of this objective is to formalize the similarity between $\textbf{r}$ and the points in the feasible set of \eqref{prob:F-main-vec}. 
Once the formalization is there, the output of the projection function will be the point with the highest similarity. In \eqref{prob:project-layer}, the distance, measured by the euclidean norm, is chosen to measure the similarity, i.e., the lower the distance, the higher the similarity. The upside of this choice is that \eqref{prob:project-layer} becomes a quadratic program, which can be solved  efficiently \cite{boyd2004convex}; thus offering a reasonable evaluation complexity once composed with $\mathcal{N}_r$. Fig. \ref{fig:projection} (left one) illustrates how \eqref{prob:project-layer} works. 
The projection function \eqref{prob:project-layer} is an instance of the DCO layer. The implementation details of this function and its integration with automatic-differentiation frameworks like PyTorch are available in \cite{diff-layers, agrawal2019differentiable}.

\subsection{Neural Network Architecture}
As depicted in \eqref{diff-proj}, the overall neural network ($\mathcal{N}_p$) is the composition of the projection function (Proj) and a backbone neural network ($\mathcal{N}_r$) as presented in Fig.~\ref{fig: NN-arc}. Since the focus of this work is on the design of the projection function, we consider a neural network with fully connected layers as the backbone.  The architecture is composed of fully connected layers with the ReLU activation function at the hidden layers. The input to the neural network is the vector form of tensor $\textbf{H}$, i.e. $\textbf{h} \in \mathbb{R}^{BQB \times 1}$, where we have $H_{i,j,k} = h_{(j-1)B^2 + (k-1)B + i}$. 


Subsequently, the input dimension is $BU$. The output will have the same dimension as the power vector, i.e., $U$. The final layer's activation function is sigmoid to bound the output of $\mathcal{N}_r$ between zero and one. Experiments showed that doing this improves the computation time of the optimization problem of the projection layer \eqref{prob:project-layer}. Since the final neural network ($\mathcal{N}_p$) uses an implicit projection, we refer to it as \textbf{D}ifferentiable \textbf{I}mplicit \textbf{P}rojection \textbf{NET}work, or for short \textbf{DIPNet}. The proposed framework is agnostic to the DNN architecture; thus, one can extend to other  DNN architectures to enhance the performance even further.

\textcolor{black} {Considering a backbone DNN (with fully-connected layers) along with proposed  projection methods enables us to provide a fair comparison with conventional PNet. The conventional PNet uses  the same backbone DNN with fully connected layers along with a softmax layer to handle the maximum transmit power constraint, but the QoS constraint was added as an extra term to the loss function along with a tuning parameter. This model is widely used in the literature (PCNet for example). Subsequently,  we are able to highlight the gains of having a powerful differentiable projection layer solely. It can be seen in the experimental evaluation part that having this layer will provide a 100\% guarantee of constraint satisfaction as compared to conventional PNet.}

\section{Differentiable Explicit Projection Framework}
In this section, we describe another way of incorporating constraints at the DNN's output. Specifically, we design a projection function via an iterative process. This idea is  inspired by \cite{donti2021dc3}, where a process called correction is applied iteratively to the output of the DNN to make it fall into the feasible set of inequality constraints. 

\subsection{Differentiable Iterative Projection}

Consider the following formulation of \eqref{prob:main:vec} or \eqref{prob:F-main-vec}, where all the constraints are concatenated to form a single vector and are denoted as $g(\textbf{p}, \textbf{H}) \leq \textbf{0}$, i.e.,

\begin{equation}
	\begin{aligned} 
		\underset{\textbf{p}}{\text{maximize}}
		& \quad R(\textbf{p},\textbf{H}) \\
		\text{subject to}
		& \quad g(\textbf{p}, \textbf{H}) \leq \textbf{0}\\
	\end{aligned}
	\label{prob:main:general}
\end{equation}
where $g: \mathbb{R}^{U} \times \mathbb{R}^{BU} \longrightarrow \mathbb{R}^{G}$ contains all the inequality constraints of the main problem ($G = UBU$). Based on \eqref{prob:main:vec}, $g$ is an affine transformation w.r.t. $\textbf{p}$. Let us define $V_{H}: \mathbb{R}^U \longrightarrow \mathbb{R}$ as a measure of the constraint violation, i.e.,
\begin{equation}
	V_{H}(\textbf{p}) = ||\mathrm{max}(g(\textbf{p}, \textbf{H}), 0)||_2^2
\end{equation}

This means that given two arbitrary points $\textbf{x}_1, \textbf{x}_2 \in \mathbb{R}^U$ if $V_{H}(\textbf{x}_1) < V_{H}(\textbf{x}_2)$, $\textbf{x}_2$ violates the constraints of \eqref{prob:main:general} more than $\textbf{x}_1$. In what follows, we define the correction process.
\begin{definition}
	An explicitly defined function $\rho: \mathbb{R}^{U} \longrightarrow \mathbb{R}^U$ that has the following properties is called correction process:
	if $\textbf{y} = \rho(\textbf{x})$, then $V_H(\textbf{y}) < V_H(\textbf{x})$, and  $\frac{\partial \textbf{y}}{\partial \textbf{x}}$ is calculable. The former condition makes sure that the output of the correction process is closer to the feasible set than its input, and the latter guarantees the differentiability of the correction process.
\end{definition}
We also denote $\rho^t$ as applying $\rho$ for $t$ times. Based on the first condition in Definition 2, by applying $\rho$ iteratively, we will end up with a point that meets the constraints i.e.,
\begin{equation}
	\textbf{p} = \mathrm{Proj}(\textbf{r}) = \lim_{t \longrightarrow \infty} \rho^t(\textbf{r}).
	\label{explicit-projection-inf}
\end{equation}
Since $\rho$ is explicitly defined and differentiable, $\rho^t$ and $\mathrm{Proj}$ are also explicitly defined and differentiable, and their Jacobian can be derived using the chain rule. Thus, the resulting projection function in \eqref{explicit-projection-inf} is explicitly defined and follows the definition of the differentiable projection function. In other words, given an input $\textbf{r} \in \mathbb{R}^U$ and output $\textbf{p}\in \mathbb{R}^U$ of the projection function, i.e., $\textbf{p} = \mathrm{Proj}(\textbf{r})$, $V_{H}(\textbf{p}) = 0$ and the Jacobian of $\textbf{p}$ w.r.t $\textbf{r}$, i.e. $\frac{\partial \textbf{p}}{\partial \textbf{r}}$ is derivable. The former condition enables end-to-end training of the whole system, i.e. letting the gradients of the loss function w.r.t. the DNN's parameters be calculated via backpropagation. 

Due to the infinite limit, the projection function as defined in \eqref{explicit-projection-inf} cannot be realized in practice. Hence, similar to \cite{donti2021dc3}, we use the truncated version of it, i.e. Proj$(.) = \rho^t(.)$ for some finite $t$. Due to truncation,  the output of the projection function may not lie on the feasible set. However, we can speed up the convergence rate of $\rho^t$ by carefully designing  $\rho$, discussed later, and choosing the initial point $\textbf{r}$. The latter can be handled by making the output of the backbone neural network $\mathcal{N}_r$  closer to the feasible set. To do this, we use the following loss function, called soft-loss, during training, i.e.,
\begin{equation}
	l_{\mathrm{soft}}(\textbf{p}, \textbf{H}) = -R(\textbf{p}, \textbf{H}) + \lambda V_H(\textbf{p}),
	\label{soft-loss}
\end{equation}
where $\textbf{p} = \rho^{t}(\mathcal{N}_r(\textbf{H};\textbf{w}_r))$, $t$ is a finite number, and $\lambda$ is a hyperparameter controlling the  constraint satisfaction relative to objective optimization. The second term is added to the loss function to penalize the points that violate the constraints, which will make $\mathcal{N}_r$ to output a good initial point to $\rho^t$. Moreover, since it is not necessary to fully satisfy the constraints during training, fewer iterations can be used to speed up the training process. During the test time, however, the number of iterations is increased to output a feasible point \cite{donti2021dc3}.

\subsection{Design of the Correction Process ($\rho$)}

Following \cite{donti2021dc3}, we use gradient-descent-based methods to realize the correction process $\rho$. Let $\nabla^{V_H}(\textbf{x}) \in \mathbb{R}^U$ and $\mathcal{H}^{V_H}(\textbf{x}) \in \mathbb{R}^{U \times U}$ denote the gradient and Hessian of $V_H$ w.r.t. $\textbf{x}$ and defined, respectively, as follows: 
\begin{align}
	\nabla^{V_H}(\textbf{x}) &= \nabla_{\textbf{x}}||\mathrm{max}(g(\textbf{x}, \textbf{H}), 0)||_2^2 \nonumber\\&= 2J^g(\textbf{x}, \textbf{H})^{\mathrm{T}}\mathrm{max}(g(\textbf{x}, \textbf{H}), 0).
	\label{grad-genral1}
\end{align}
\begin{align}
	&\mathcal{H}^{V_H}(\textbf{x}) = 2I(\mathrm{max}(g(\textbf{x}, \textbf{H}), 0))^{\mathrm{T}}\textbf{K}^g(\textbf{x}, \textbf{H}) + \nonumber\\& 2\mathrm{max}(g(\textbf{x}, \textbf{H}), 0)^{\mathrm{T}}\textbf{T}^g(\textbf{x}, \textbf{H}).
	\label{hessian-general1}
\end{align}
where $J^g(\textbf{x}, \textbf{H}) \in \mathbb{R}^{G \times U}$ is the Jacobian of $g$ w.r.t. $\textbf{x}$. 
Moreover, $I: \mathbb{R} \longrightarrow \mathbb{R}$, $\textbf{K} \in \mathbb{R}^{ G\times U \times U}$, and $\textbf{T} \in \mathbb{R}^{G \times U \times U}$ are defined as shown in \eqref{top}. 
\begin{figure*}
	\begin{equation}
		\label{top}
		I(x) = \left\{
		\begin{array}{ll}
			1  & \mbox{if } x > 0 \\
			0 & \mbox{otherwise}
		\end{array} \right., \quad
		K^g(\textbf{x}, \textbf{H})_{k, i, j} = \frac{\partial g_k}{\partial x_i}\frac{\partial g_k}{\partial x_j}, \quad T^g(\textbf{x}, \textbf{H})_{k, i, j} = \frac{\partial^2 g_k}{\partial x_i \partial x
			_j}.
	\end{equation}
	\hrule
\end{figure*}
where $I$ is an indicator function that is applied element-wise to $\mathrm{max}(g(\textbf{x}, \textbf{H}), 0)$, $\textbf{K}$, and $\textbf{T}$ are rank-3 tensors. The definition of Hessian in \eqref{hessian-general1} contains vector to tensor dot product, resulting in a $U \times U$ matrix. The product is computed as follows:
\begin{equation}
	\textbf{C}= \textbf{a}^{\mathrm{T}}\textbf{Z} \longrightarrow C_{i,j} = \sum_{k = 1}^{G} a_k Z_{k,i,j},
\end{equation}
where $\textbf{Z} \in \mathbb{R}^{G \times U \times U }$,  $\textbf{a} \in \mathbb{R}^{G \times 1}$, and $\textbf{C} \in \mathbb{R}^{U \times U}$. 

The feasible set of \eqref{prob:main} can be presented with linear constraints, i.e., a polyhedron, as can be seen in \eqref{prob:main:vec}. Thus, in the following, we derive the above-mentioned formulas in case $g$ is an affine function, i.e. $g(\textbf{x}, \textbf{H}) = \textbf{M}\textbf{x} + \textbf{n}$, where $\textbf{M}: \mathbb{R}^{G \times U}$ and $\textbf{n} \in \mathbb{R}^{G \times 1}$ are functions of $\textbf{H}$.
\begin{equation}
	g(\textbf{x}, \textbf{H}) = \textbf{M}\textbf{x} + \textbf{n} \longrightarrow 
	\left\{
	\begin{array}{ll}
		J^g(\textbf{x}, \textbf{H}) = \textbf{M} \\
		K^g(\textbf{x}, \textbf{H})_{k, i, j} = M_{k,i}M_{k,j} \\
		T^g(\textbf{x}, \textbf{H})_{k, i, j} = 0
	\end{array} \right.
\end{equation}
Unlike DIPNet, DEPNet does not have a convexity requirement. 

Now that we have access to the first and second-order information of $V_H$ w.r.t. $\textbf{x}$, the correction process can be formulated as follows:
\begin{equation}
	\rho(\textbf{x}) = \textbf{x} + \Delta \textbf{x},
\end{equation}
where $\Delta \textbf{x}$ can take the form of variations of descent methods. Examples of which include vanilla gradient descent ($\Delta \textbf{x} = -\gamma \nabla^{V_H}(\textbf{x})$) or Newton method ($\Delta \textbf{x}= -\mathcal{H}^{V_H}(\textbf{x})^{-1} \nabla^{V_H}(\textbf{x})$). Here, we used gradient-descent with momentum \cite{goodfellow2016deep} during training and the Newton method during test for faster training and lower violation probability at test. In the following, the correction process formulation for these choices is provided.

Let $\Delta \textbf{x}^{t}$ be the step of the correction process that is applied at the iteration $t$ of the projection function. The update rule for gradient descent with momentum can be given as follows:
\begin{equation}
	\Delta \textbf{x}^{t} = - \gamma\nabla^{V_H}(\textbf{x}) - \mu  \Delta \textbf{x}^{t-1},
	\label{grad-moment}
\end{equation}
where $\gamma$ is called step-size or learning rate and $\mu$ is called the momentum. To avoid confusion with the learning rate involved in training DNNs, we refer to $\gamma$ as the step-size. It should be noted that $\gamma$ is chosen prior to training and is fixed during the training.  This is to make the execution time of the correction process faster and the calculation of its derivatives easier during the training. In this case, having momentum will help the convergence of gradient descent by making it less vulnerable to the oscillations of noisy gradients \cite{goodfellow2016deep}.


One can use alternatives like exact or  backtracking line-search \cite{boyd2004convex} for choosing $\gamma$ in each iteration. Although it helps with convergence, it introduces more computational complexity during the training, and the calculation of the derivative of the correction process will not be as straightforward as using a fixed value for the step-size. At the test time, however, we can use these techniques to fasten the convergence rate of the correction process and get feasible solutions from the projection function. Inspired by this idea, instead of using gradient descent at the test time, we use Newton method, which has a faster convergence rate \cite{boyd2004convex}. Since $\mathrm{max(-,0)}$ is used in the definition of $V_{H}(.)$ and its second derivative is zero, the hessian matrix is poorly conditioned and has lots of zeros, making it not invertible. To deal with this issue, we regularize the hessian matrix by adding a small value $\alpha$ to its diagonal. Thus, the update rule becomes: 
\begin{equation}
	\Delta \textbf{x}^{t}= -(\mathcal{H}^{V_H}(\textbf{x}) + \alpha \textbf{I})^{-1} \nabla^{V_H}(\textbf{x})
	\label{newton}
\end{equation}
where $\textbf{I} \in \mathbb{R}^{U \times U}$ is the identity matrix. In the following, for simplicity, we refer to \eqref{newton} as the Newton method update rule. \eqref{newton} uses the second-order information of the objective at hand. As proven in \cite{boyd2004convex} and shown in my experiments, Newton method has a faster convergence rate than gradient-descent. The only downside of it is that the calculation of the derivative of its steps is not straightforward due to matrix inversion in \eqref{newton}. Thus, we only used it at test time, and utilized gradient descent during training. {Moreover, to make sure that the values of $\textbf{x}$ are always non-negative, which is very important for non-convex cases, we apply $\mathrm{max}(\textbf{x}, 0)$ in an element-wise manner on $\textbf{x}$ after each update of the correction process.}
\subsection{Neural Network Architecture}
For consistency and the sake of fair comparison,  we use the same DNN architecture, as described in Section~IV, with minor modifications. As depicted in Fig.~\ref{fig: NN-arc}, after the last affine transformation, there is a sigmoid non-linearity followed by a projection function. Using sigmoid at the final layer of $\mathcal{N}_r$ showed faster convergence of this projection method in the experiments. Since this model uses an explicitly defined differentiable transformation to realize the projection, we call it \textbf{D}ifferentiable \textbf{E}xplicit \textbf{P}rojection \textbf{NET}work (DEPNet).

\section{An enhancement: Frank-Wolfe Algorithm}
The optimality of the output of the projection function is dependent on the quality of the initial point ($\textbf{r}$) and the projection method. Although the quality of the initial point will improve during the training, it might show sub-optimality. This means that the final power profile is feasible, but might not be optimal. A remedy is to apply constrained optimization algorithms, where the output of the projection function  is passed as the initial point of the chosen algorithm ($\textbf{p}^0$). The algorithm, then, outputs an enhanced power profile that achieves a higher sum-rate than $\textbf{p}^0$.

Among several variants of general constraint optimization algorithms, we select Frank-Wolfe \cite{lacoste2016convergence}, a.k.a conditional gradient descent, due to its low computational cost  in each iteration of the algorithm. Compared to other alternatives like projected gradient descent which requires solving a quadratic program in each iteration, Frank-Wolfe only solves a linear program in each iteration \cite{lacoste2016convergence}. In the following, the description of the Frank-Wolfe algorithm is provided.

As mentioned earlier, the initial point of this algorithm will be the output of the projection function, i.e., $\textbf{p}^0 = \mathrm{Proj}(\textbf{r})$. The algorithm is agnostic to the projection method and only requires a feasible initial point. Thus, it can be used with both DIPNet and DEPNet. The final power profile will be the output of this algorithm.  Let us denote the output of the Frank-Wolfe algorithm after $t$ and $t + 1$ iterations as $\textbf{p}^{t}$ and $\textbf{p}^{t + 1} \in \mathbb{R}^U$, respectively. They belong to the feasible set of \eqref{prob:main:vec} and the relation between them is:
\begin{equation}
	\textbf{p}^{t + 1} = \textbf{p}^t + \lambda \textbf{p}^{t + \frac{1}{2}}, \quad \lambda \in [0,1]
	\label{step-fw}
\end{equation}
where $\textbf{p}^{t + \frac{1}{2}}$ is an intermediate variable derived as follows:
\begin{equation}
	\begin{aligned} 
		\textbf{p}^{t+\frac{1}{2}} = \underset{\hat{\textbf{p}}}{\text{argmin}}
		& \quad <\hat{\textbf{p}}, \nabla_{p}R(\textbf{p}^t,\textbf{H})> \\
		\text{subject to}
		& \qquad \hat{\textbf{p}} \geq \textbf{0}, \quad \textbf{A}\hat{\textbf{p}} \leq P_{\mathrm{max}}\textbf{1}, \quad \textbf{C}\hat{\textbf{p}} \geq \textbf{d}\\
	\end{aligned}
	\label{frank-wolfe}
\end{equation}
where the objective is the euclidean dot product of the optimization variable and the gradient of sum-rate ($R(\textbf{p},\textbf{H})$) w.r.t  $\textbf{p}^t$. Since \eqref{frank-wolfe} is solved over the feasible set of \eqref{prob:main:vec}, $\textbf{p}^{t + \frac{1}{2}}$ belongs to the feasible set of \eqref{prob:main:vec} as well. Following \eqref{step-fw}, a convex combination of $\textbf{p}^{t + \frac{1}{2}}$ and $\textbf{p}^t$ is chosen as the output of the $t$ + 1 iteration. Since the feasible set of \eqref{prob:main:vec} is convex, $\textbf{p}^{t + 1}$ will be a feasible power profile. $\lambda$ is the step size that can be chosen via line search or adaptive methods \cite{lacoste2016convergence}.

\section{Experimental Set-up and Benchmarks}
In this section, we describe the dataset generation procedure, system set-up, and relevant benchmarks to evaluate the performance of the proposed DIPNet and DEPNet \footnote{The codes to reproduce the simulation results are available on https://github.com/Mehrazin/Power-Control-with-QoS-Guarantees}.

\begin{table*}[t]
	\begin{center}
		\caption{A comparison of  DIPNet, DEPNet, and optimization benchmark in terms of sum-rate (in Mbps) and computation time (in msec).}
		\label{table:more}
		\begin{tabular}{|l|llll|ll|ll|ll|}
			\hline
			\multirow{2}{*}{\textbf{Dataset ID}} & \multicolumn{4}{l|}{\textbf{Configuration}}                                               & \multicolumn{2}{l|}{\textbf{DIPNet}}               & \multicolumn{2}{l|}{\textbf{DEPNet}}               & \multicolumn{2}{l|}{\textbf{GP}}                   \\ \cline{2-11} 
			& \multicolumn{1}{l|}{BSs} & \multicolumn{1}{l|}{Users} & \multicolumn{1}{l|}{Quota} & Target Rate & \multicolumn{1}{l|}{Sum-rate} & Time & \multicolumn{1}{l|}{Sum-rate} & Time & \multicolumn{1}{l|}{Sum-rate} & Time \\ \hline
			1                            & \multicolumn{1}{l|}{5} & \multicolumn{1}{l|}{5} & \multicolumn{1}{l|}{1} & 10     & \multicolumn{1}{l|}{104.15}         &      2.91     & \multicolumn{1}{l|}{107.35}         &      0.69     & \multicolumn{1}{l|}{119.6}         &      478.47     \\ \hline
			2                            & \multicolumn{1}{l|}{2} & \multicolumn{1}{l|}{6} & \multicolumn{1}{l|}{3} & 2.5   & \multicolumn{1}{l|}{243.35}         &   2.70        & \multicolumn{1}{l|}{244.6}         &       0.46    & \multicolumn{1}{l|}{244}    & 224.22         \\ \hline
			3     & \multicolumn{1}{l|}{4}  & \multicolumn{1}{l|}{20}  & \multicolumn{1}{l|}{5}  & 5      & \multicolumn{1}{l|}{510.05}         &     4.06      & \multicolumn{1}{l|}{515.2}         &   3.06        & \multicolumn{1}{l|}{569.6}         &       7852.19    \\ \hline
			4       & \multicolumn{1}{l|}{4}  & \multicolumn{1}{l|}{20}  & \multicolumn{1}{l|}{5}  &   2.5    & \multicolumn{1}{l|}{456.1}         &     4.02      & \multicolumn{1}{l|}{458.95}         &     3.05      & \multicolumn{1}{l|}{512.9}         &     7864.58     \\ \hline  5 & \multicolumn{1}{l|}{6}  & \multicolumn{1}{l|}{24}  & \multicolumn{1}{l|}{4}  &    5   & \multicolumn{1}{l|}{441.25}         &     5.85      & \multicolumn{1}{l|}{444.2}         &     4.99      & \multicolumn{1}{l|}{515.4}         &       35607.32   \\ \hline 6 & \multicolumn{1}{l|}{6}  & \multicolumn{1}{l|}{24}  & \multicolumn{1}{l|}{4}  &  2.5     & \multicolumn{1}{l|}{397.7}         &   6.05        & \multicolumn{1}{l|}{399.85}         &    5.81       & \multicolumn{1}{l|}{466.55}         &     36338.41      \\ \hline 7 & \multicolumn{1}{l|}{5}  & \multicolumn{1}{l|}{5}  & \multicolumn{1}{l|}{1}  &  2.5     & \multicolumn{1}{l|}{55.7}         &    2.92       & \multicolumn{1}{l|}{57.4}         &      0.66     & \multicolumn{1}{l|}{68.35}         &    530.52       \\ \hline 8 & \multicolumn{1}{l|}{4}  & \multicolumn{1}{l|}{12}  & \multicolumn{1}{l|}{3}  & 2.5      & \multicolumn{1}{l|}{258.7}         &   3.82        & \multicolumn{1}{l|}{261.9}         &  1.73         & \multicolumn{1}{l|}{297}         &     2244.52      \\ \hline
		\end{tabular}
	\end{center}
\end{table*}

\subsubsection{DataSet Generation}
For the experimental evaluation, we used two datasets. The first one, called \textit{Gaussian Dataset}, follows the symmetric interference channel model with independent and identically distributed (i.i.d.) Rayleigh fading for all channel coefficients.
The maximum power is set to 1 mW, the transmission bandwidth is 5 MHz, and the thermal noise power is assumed to be 0.01$\mu W$ \cite{4801507}. This model is widely adopted in the literature to evaluate resource allocation algorithms\cite{liang2019towards, 4801507, sun2018learning}. The second dataset, called \textit{Path-loss Dataset}, has channel coefficients that are composed of large-scale path-loss, shadowing, and fading. The dataset generation has the following steps. First, the locations of the BSs and users are determined by random sampling from a 500 m $\times 500$~m area. We consider that the distance between the two closest BSs, BSs, and user, and between closest users, should be at least 100 m, 5 m, and 2 m, respectively. 
The carrier frequency is 2.4 GHz, the transmission bandwidth is 5 MHz, and the noise spectral density is set to -169 dBm. The remaining details of the channel model can be found in \cite{shi2014group}. 

For both datasets, once a datapoint or channel realization is generated, we associate the best $Q$ users to the first BS in terms of their channels. These users are then excluded from the user set. Then, the users' of the next BS are determined by the same process. This process is continued until all the users are associated with a BS. After that, for each BS, the allocated set of users is sorted out based on their channel gains and then gets assigned to the channels such that the first channel of each BS has the strongest user of that BS. 

Different datasets are generated under different configurations.  For each configuration, 100,000 data samples are generated and divided to train, validation, and test with 0.9, 0.05, and 0.05 ratios, respectively. Unless stated otherwise, the number of BSs is set to $B=$4. 
Also, we set the minimum users' rate requirement  to $\alpha=$ 2.5 Mbps unless stated otherwise. 
To provide a comprehensive quantitative analysis, additional datasets are generated under different configurations listed in Table~II.  Dataset ID is used to refer to different configurations, where the quota of each BS is given as $Q=U/B$. All datasets in Table~II are following the Path-loss models. The  number of BSs, users, and the minimum rate are different among them. 

\subsubsection{Feasibility Check}
In the dataset preparation process, we need to make sure that all data points are feasible, i.e., there exists a power profile that meets the constraints of problem \eqref{prob:main}. For that, we use the approach mentioned in \cite{chiang2008power} and used in \cite{4801507,liang2019towards}. The  approach provides feasible transmit powers as well as minimum transmit powers that can fulfill the minimum rate constraints. Originally, this approach is not designed for the case where there is more than one channel, i.e., $Q > 1$. However, since channels on a given BS are orthogonal, we  can break \eqref{prob:main} into $Q$ sub-problems, and apply the feasibility check approach with slight modifications. The description of the approach is defined next. 
Let us define matrix $\textbf{B}^q$ as follows:
\begin{equation}
	B^q_{b,\hat{b}} = 
	\left\{
	\begin{array}{ll}
		0  &  b = \hat{b} \\
		\frac{\beta_{b,q}|h_{b,q, \hat{b}}|^2}{|h_{b,q,b}|^2} & b \neq \hat{b}
	\end{array} \right., \quad
\end{equation}
If the maximum eigenvalue of $\textbf{B}^q$ is larger than 1, there is no feasible solution. That is, we can not find a power vector for channel $q$ that can satisfy the minimum rate requirement of the users associated with this channel. Otherwise, we can find a feasible power allocation profile as:
\begin{equation}
	\textbf{P}_{:,q} = (\textbf{I} - \textbf{B}^q)^{-1}\textbf{u}^q,
	\label{min-power}
\end{equation}
where $\textbf{P}_{:,q}$ is the transmit power vector over channel $q$, $\textbf{I}$ is a $B\times B$ identity matrix, and $\textbf{u}^q$ is a $B\times1$ vector with $j$th element given as
$
u^q_b = \frac{\beta_{b,q}\sigma^2}{|h_{b,q,b}|^2}.
$
Once all  power vectors are calculated for $q \in \mathcal{Q}$, we create the final power matrix in the following manner.
If $\textbf{P}$ meets the first and second constraints of \eqref{prob:main}, i.e., all elements of \textbf{P} are greater than zero and the sum of each row is lesser than $P_{\mathrm{max}}$, $\textbf{P}$ is a feasible solution of \eqref{prob:main}. Otherwise, \eqref{prob:main} is not feasible.

\subsubsection{Benchmarks}
We consider three main benchmarks\footnote{\textcolor{black} {Traditionally greedy algorithms are considered as a common benchmark. However, in the presence of QoS constraint, the greedy algorithms are not suitable and can not offer us a fair comparison with the proposed scheme. Thus, similar to the relevant research works, such as \cite{liang2019towards}, geometric programming (GP) \cite{chiang2008power} is considered as the main optimization-based benchmark.}}, i.e.,
\begin{itemize}
	\item \textbf{PNet:} is a neural network model exactly like DIPNet and DEPNet, but without the projection layer, i.e. FCNN. This model is an extension of PCNet \cite{liang2019towards} that works for the multi-channel scenario. The power budget constraint was handled by having softmax as the activation function of the final layer. The violation of the QoS constraint was added as a penalty term to the loss function, similar to the soft loss as in Section~VI. This  DNN-based benchmark  does not utilize the proposed projection methods.
	\item \textbf{GP:} is an optimization-based solution that uses the  high-SINR assumption to transform the main problem \eqref{prob:main} to a geometric program \cite{chiang2008power}. 
	{\item \textbf{Genetic Algorithm:} is a well-known global optimization algorithm with high complexity.}

\end{itemize}

\section{Numerical Results and Discussions}

In this section, we present the performance of the proposed methods (DIPNet and DEPNet) with the conventional benchmarks. The considered performance metrics include network sum-rate, QoS violation probability, and per sample test time. To get a good estimate of the computation time, the overall computation time of each model is measured over the test set. Per-sample computation time is then calculated by averaging over the data points. For the calculation of the sum rate, if the output of a scheme violates the QoS constraint, we set the sum rate of that datapoint to be zero. In this way, the effect of constraint violation is highlighted in the performance evaluation. This method is adopted from \cite{9281322}.

\begin{figure*}[!ht]
	\begin{minipage}{0.5\textwidth}
		\includegraphics[scale=0.55,  trim=4 4 4 4 ,clip]{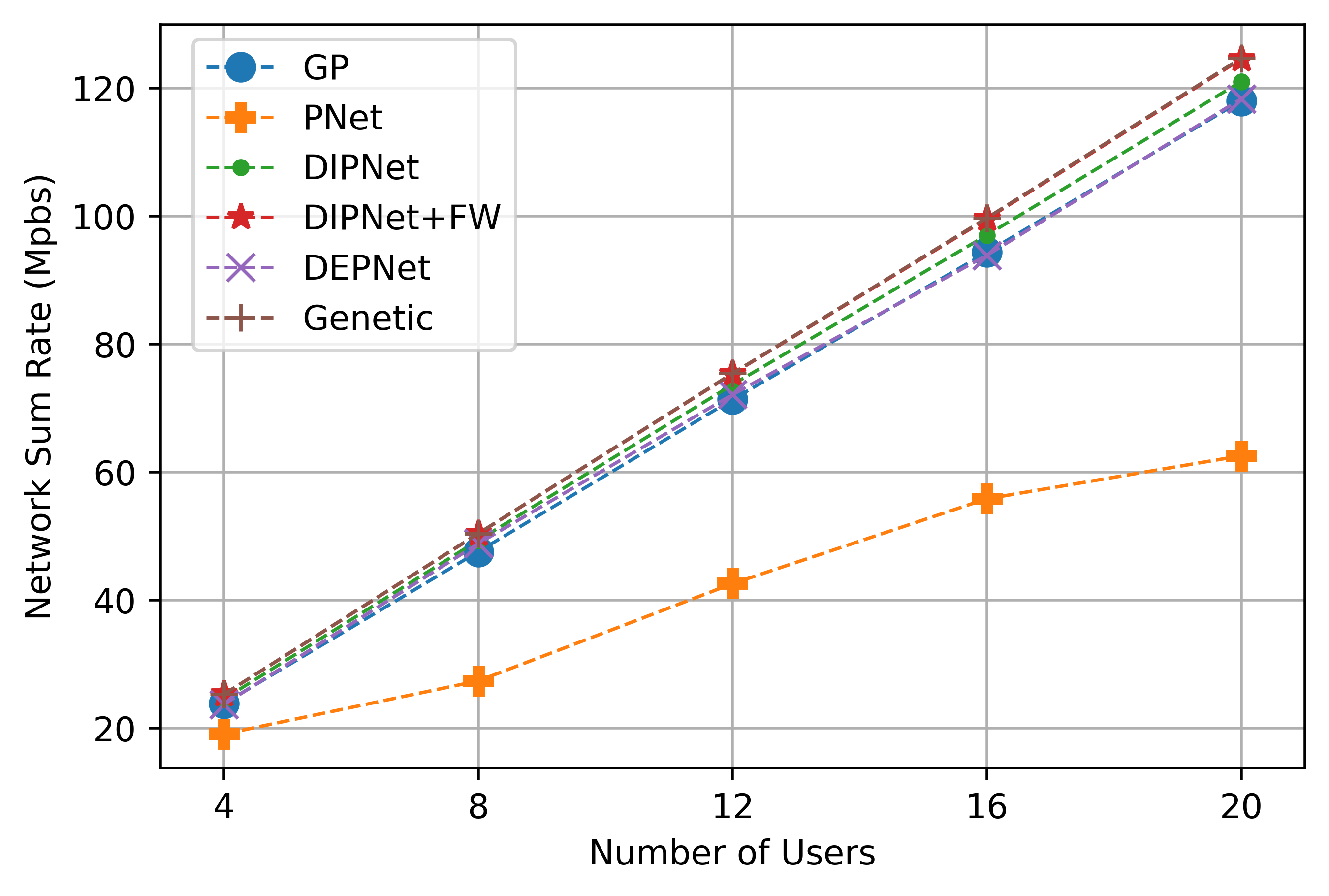}
	\end{minipage}
	\begin{minipage}{0.5\textwidth}
		\includegraphics[scale=0.55,  trim=4 4 4 4 ,clip]{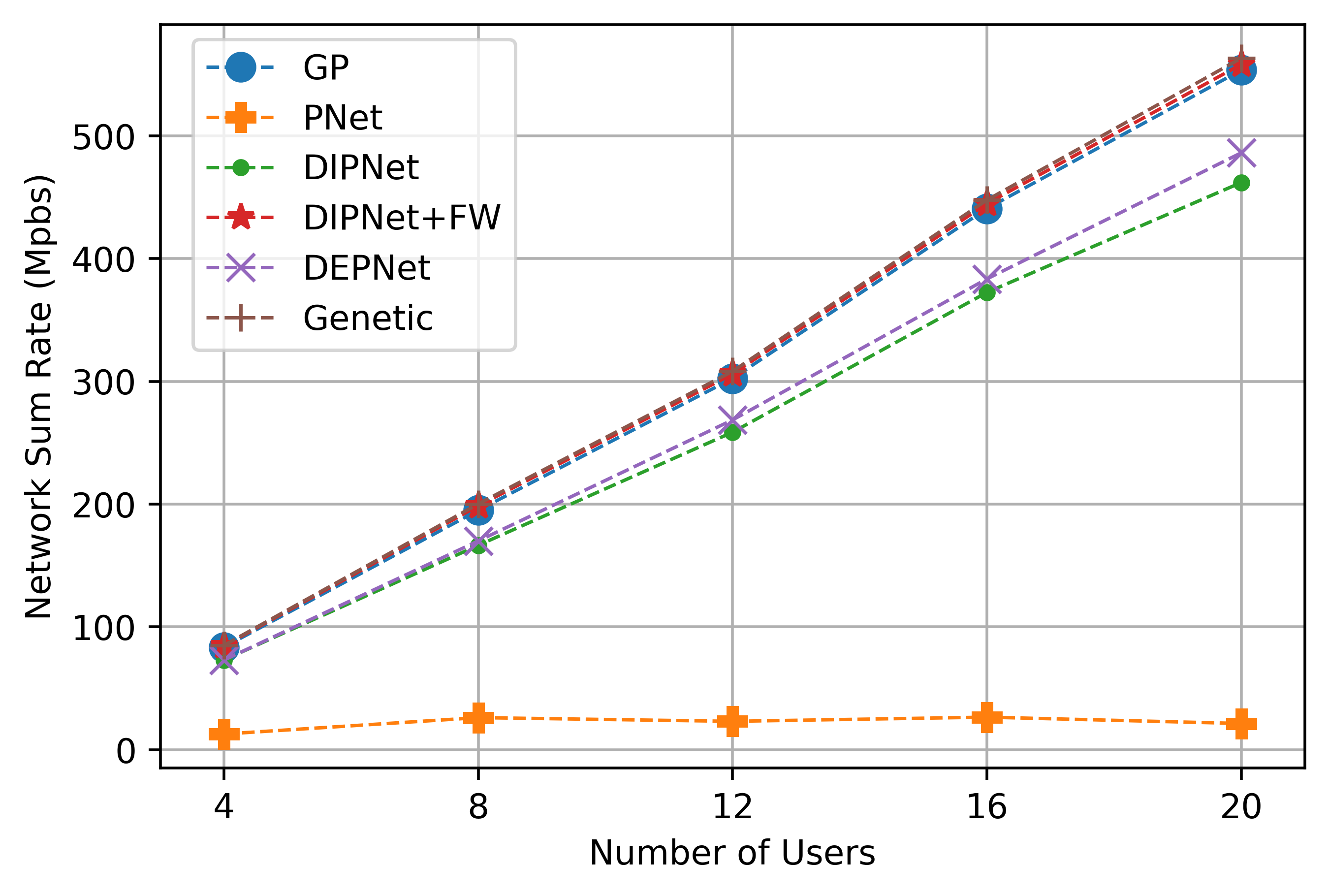}
	\end{minipage}
	\caption{Sum-rate for GP, PNet, DIPNet, DIPNet+FW, DEPNet, and Genetic (Left: Gaussian - Right: Path-loss).}
	\label{SR}
\end{figure*}

\begin{figure}[t]
	\begin{minipage}{0.25\textwidth}
		\includegraphics[scale=0.32,  trim=4 4 4 4 ,clip]{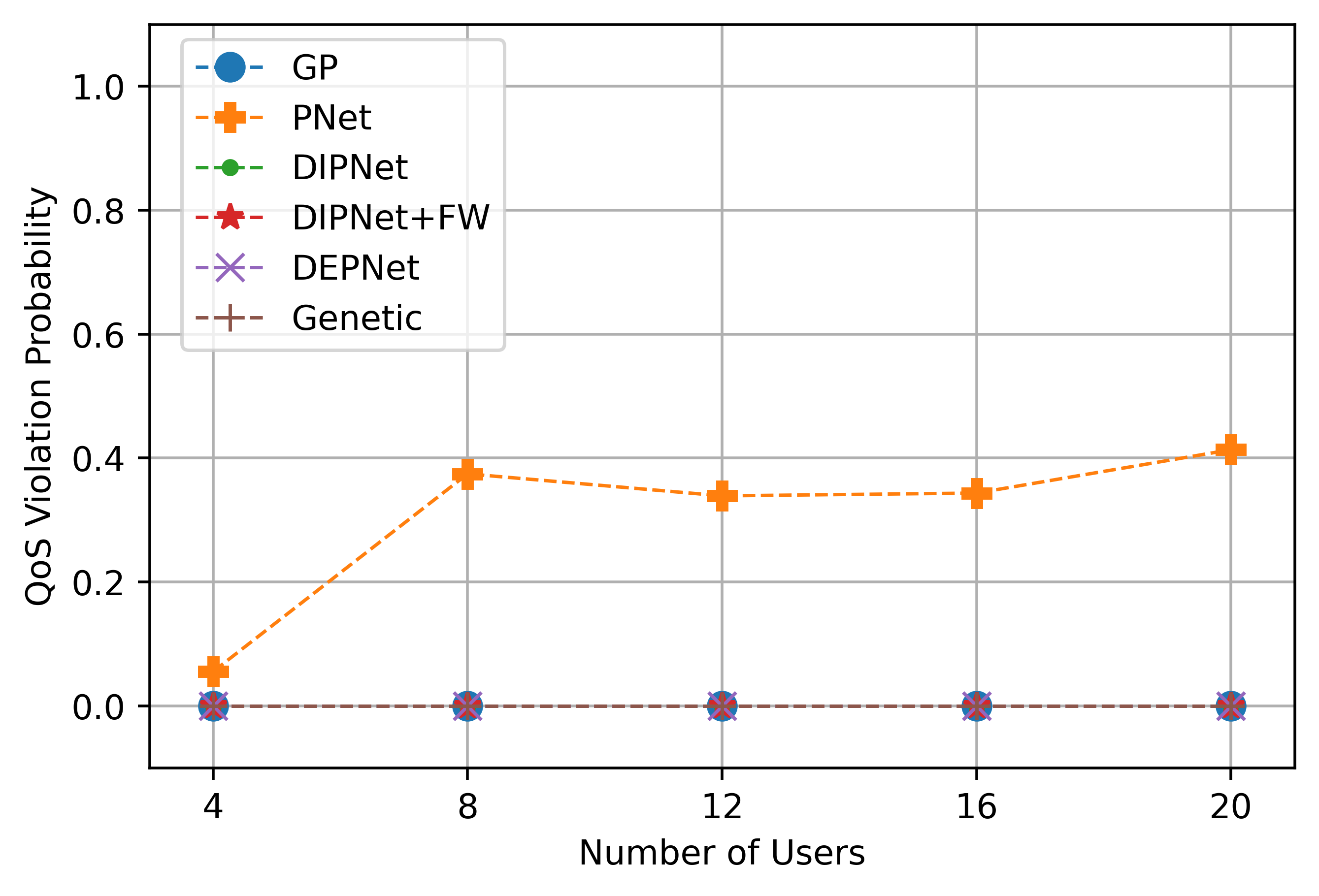}
	\end{minipage}\hspace{-2mm}
	\begin{minipage}{0.2\textwidth}
		\includegraphics[scale=0.32,  trim=4 4 4 4 ,clip]{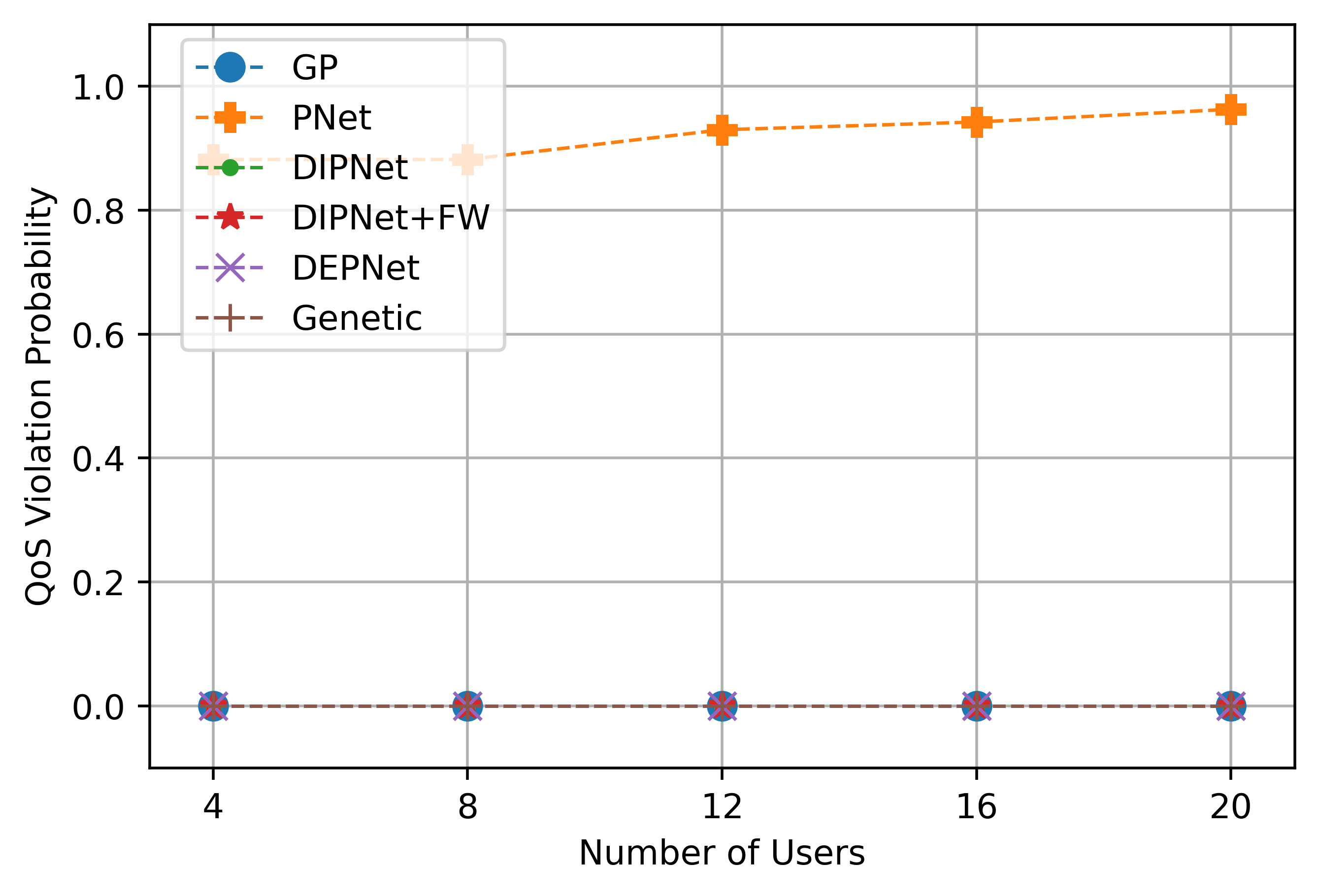}
	\end{minipage}
	\caption{QoS violation  for GP, PNet, DIPNet, DIPNet+FW, DEPNet, and Genetic (Left: Gaussian Dataset - Right: Pathloss Dataset).}
	\label{vio}
\end{figure}

\begin{figure*}[t]
	\begin{minipage}{0.5\textwidth}
		\centering
		\includegraphics[scale=0.5,  trim=4 4 4 4 ,clip]{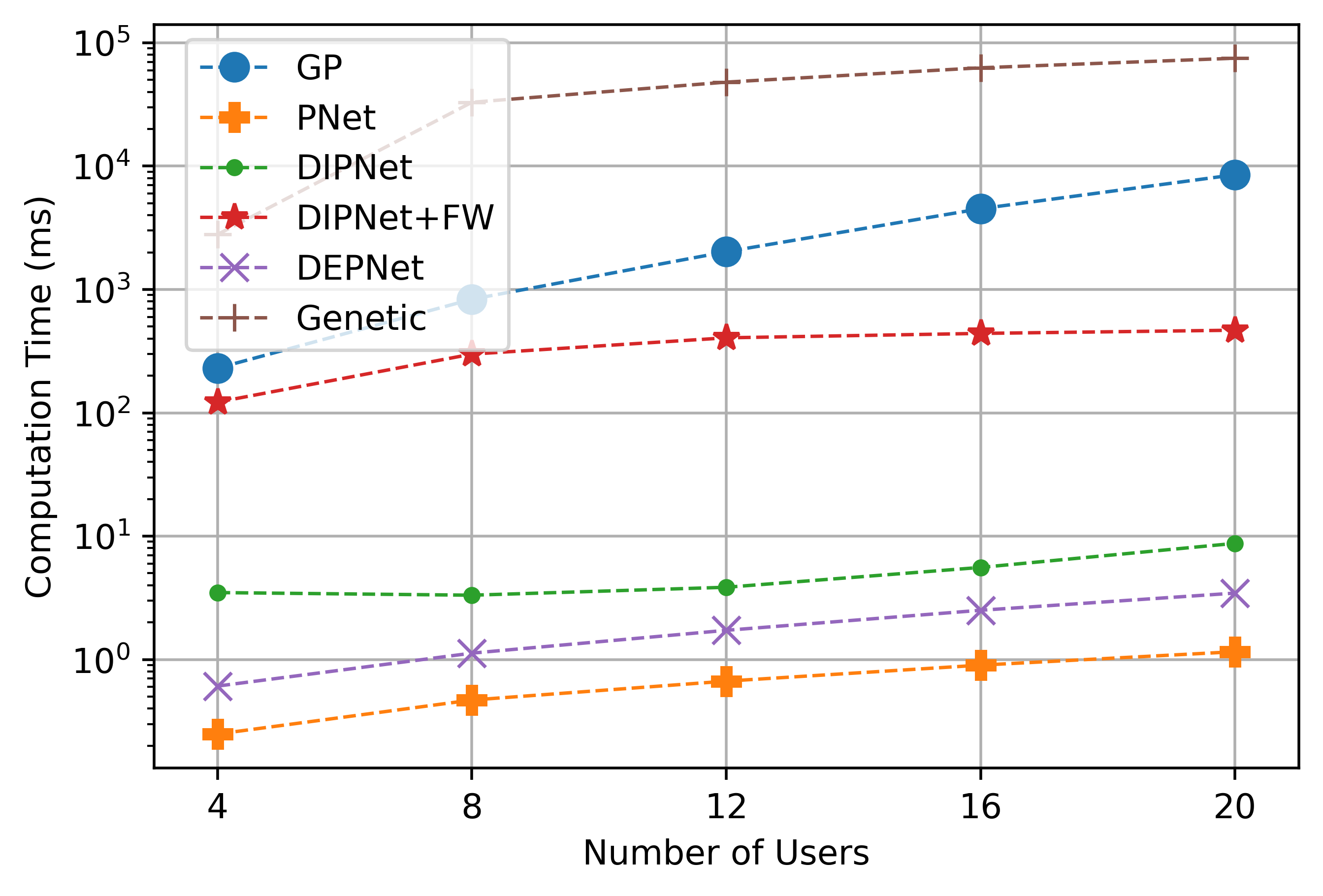}
	\end{minipage}
	\begin{minipage}{0.5\textwidth}
		\centering
		\includegraphics[scale=0.5,  trim=4 4 4 4 ,clip]{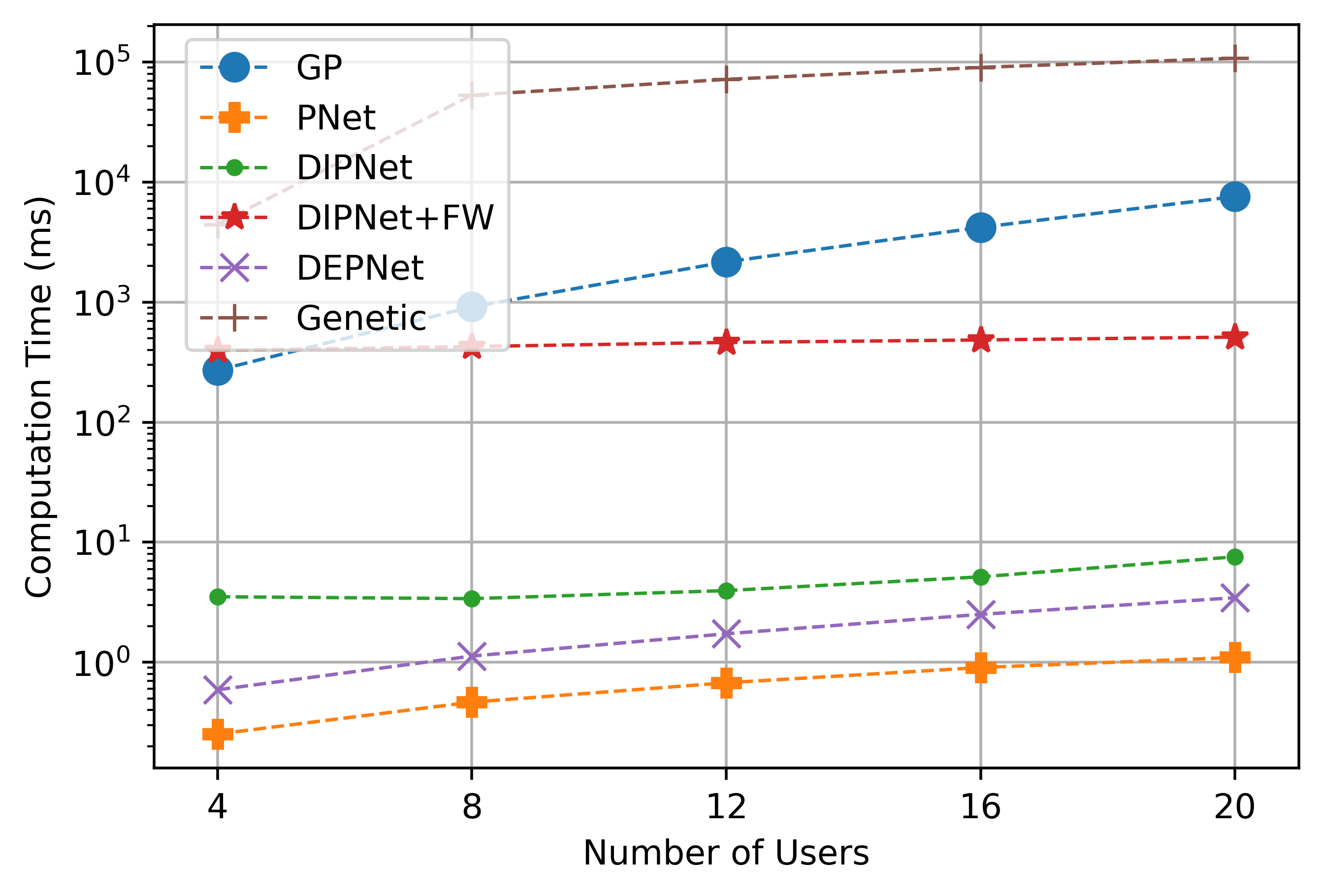}
	\end{minipage}
	\caption{Computation time for GP, PNet, DIPNet, DIPNet+FW, DEPNet, and Genetic (Left: Gaussian Dataset - Right: Pathloss Dataset).}
	\label{time}
\end{figure*}

In all the experiments, an FCNN with three 200-dimensional hidden layers is used as the backbone of DIPNet and DEPNet ($\mathcal{N}_r$). Batch normalization \cite{ioffe2015batch} and Dropout \cite{hinton2012improving} are used to accelerate the training and prevent over-fitting. The same neural network is used for the PNet as well. For the correction process of DEPNet, we used gradient descent with momentum for the training and Newton method for the testing. The momentum is set to 0.5 for all the datasets, and the step size is fine-tuned for each dataset individually from the interval of 0.5 to 0.0005. Similarly, the parameter $\lambda$ in soft-loss \eqref{soft-loss} is chosen from the interval of 10 to 10000 for each dataset individually. The number of iterations in the iterative projection is set to 5 and 100 for the training (with Gradient Descent method) and testing (with Newton method), respectively.   

For all DNNs, we used a learning rate of 0.001, batch size of 10, and learning rate decay rate of 0.99. ADAM is also used for training the DNNs. All DNNs are trained for 20 epochs, and early stopping is used to pick the model with the best performance. 
That is, after each epoch, we check the performance of the model over the validation set based on a metric like network sum-rate. Then, we pick the model that has the best performance among all epochs.
The performance metric for DIPNet and DEPNet is the network sum-rate.
For the PNet, we pick the model that has the minimum QoS violation probability. This is because choosing network sum-rate as the selecting criteria will result in a model that has a significant QoS violation probability, which will not be comparable with DIPNet and DEPNet. Moreover, based on the experiments, $\lambda$ in the soft-loss of PNet is set to 1000 for Gaussian datasets and 10000 for path-loss datasets. This results in a model that has comparable performance in terms of network sum-rate and QoS violation with DIPNet and DEPNet.

For the implementation, we used PyTorch \cite{paszke2019pytorch} as the automatic differentiation engine. For implementing the implicit projection in DIPNet, we used the CVXPYlayer package \cite{agrawal2019differentiable}. ECOS \cite{ecos} is chosen as the optimization solver of this layer among the available options.  To make the implementation consistent, GP is implemented in Python using CVXPY package \cite{diamond2016cvxpy}. MOSEK \cite{mosek}, a commercial solver, is used as the backbone solver of CVXPY for both GP and Frank-Wolfe. The maximum number of iterations for Frank-Wolfe is set to 50 and the threshold is set to 0.001. Since the output of DIPNet is always feasible, we only applied Frank-Wolfe to the output of DIPNet, referred to as DIPNet+FW. Moreover, we used the genetic algorithm's implementation on MATLAB with 5000 iterations for all the datasets. The experiments are done on a desktop computer with Intel Core i7-8700 CPU 3.20GHz and 8GB of RAM.

\subsection{Performance of DIPNet and DEPNet}
\subsubsection{Optimization-based Benchmark}
Starting with the Gaussian datasets, Fig.~\ref{SR} (left)  demonstrates the performance of DIPNet and DEPNet in terms of the achievable aggregate network sum-rate as compared to the conventional GP-based optimization solution. Both DIPNet and DEPNet demonstrate a close sum-rate to GP  while showing zero QoS violation probability (as shown in Fig.~\ref{vio} (left)). By increasing the problem size, the required computation time of GP increases drastically (as shown in Fig.~\ref{time} (left)). DIPNet and DEPNet, on the other hand, have much reduced computational complexity. 

Considering the Path-loss datasets, Fig.\ref{SR} (right) shows that GP outperforms both DIPNet and DEPNet in terms of network sum-rate. The difference in network sum-rates starts to grow as the problem size increases.  The solutions of GP are always feasible. The same is true for DIPNet and DEPNet (as shown in Fig. \ref{vio} (right)).  The time complexity of GP increases exponentially when the problem size increases (as shown in Fig. \ref{time} (right)). Once the Frank-Wolfe algorithm is applied to DIPNet results, we can observe an apparent boost in the network sum-rate for both Gaussian and Path-loss datasets (as can be seen in Fig. \ref{SR}). Compared to GP, DIPNet with Frank-Wolfe enhancement surpasses GP across all the datasets. The cost of this boost is an increase in computation time compared to DIPNet (as can be seen in Fig. \ref{time}). Although the computation time of Frank-Wolfe-based DIPNet is higher than other DNN-based methods, it is still lower than GP, especially when the number of users increases (as  in Fig. \ref{time}).

Compared with the Genetic algorithm, we can observe that the results of the GP are very close to the true optimal (Fig. \ref{SR}), which makes it a good benchmark for comparison. Also, we can see that the results of DIPNet+FW are almost similar to the genetic algorithm, with noticeably lower computational complexity (Fig. \ref{time}). Finally, Table~\ref{table:more} provides more experimental results of the DIPNet, DEPNet, and GP across various network configurations. Similar to Fig \ref{SR}, GP also, outperforms DEPNet and DIPNet in terms of network sum-rate. The difference becomes less significant when the minimum rate and the problem size is small. The main reason behind the better performance of GP is that the proposed projection methods tend to find points at the boundary of the feasible set. GP, on the other hand, can search within the feasible set to find solutions with a higher network sum-rate.

\subsubsection{A Comparison to Conventional PNet (Enhanced PCNet)}
As shown in Fig. \ref{SR}, PNet always outputs a solution that achieves a lower sum-rate than DIPNet and DEPNet for both Gaussian and Path-loss datasets. As pointed out in \cite{liang2019towards}, reducing the $\lambda$ in the soft-loss results in an increase in the constraint violation which will also reduce the resulting network sum rate of the PNet.
In Fig \ref{vio}, we note that the violation probability of PNet increases with the dimension of the problem (number of users). Comparing Path-loss and Gaussian datasets, we can see that the QoS violation probability of PNet increases dramatically while working on a more realistic dataset, i.e. Path-loss dataset. This is due to the fact that there is no mechanism in the architecture of PNet to satisfy the QoS constraints. Since PNet doesn't utilize any projection functionality, the computation time of PNet is lesser than DIPNet and DEPNet (as can be seen in Fig. \ref{time}). 
Considering all these factors, we can conclude that PNet is the best option when there is no complex hard constraint like QoS. However, once the QoS constraint is introduced, a projection function has to be utilized to satisfy the constraints with zero violation probability.


\subsubsection{DIPNet vs DEPNet}
The main difference between DIPNet and DEPNet is the projection function used in them to satisfy the QoS constraint. DIPNet uses an optimization solver to incorporate the constraint, and it is easier to implement and does not require tuning some hyperparameters for the projection function. DEPNet, however, uses an iterative process to realize the projection functionality. 
We used gradient descent with momentum during the training with five iterations. This choice is computationally efficient and leads to fast training. However, the step-size of gradient-descent needs to be tuned for each dataset; thus requiring experimentation. At the test time, we used Newton method, which has a faster convergence rate than gradient descent, with 100 iterations to make sure the output is always feasible. As shown in Fig.~\ref{vio}, both DIPNet and DEPNet achieve zero violation probability, but DEPNet is more computationally efficient than DIPNet (as can be noted from Fig.~\ref{time}).

\begin{figure*}[t]
	\centering
	\begin{minipage}{0.5\textwidth}
		\centering
		\includegraphics[scale=0.5,  trim=4 4 4 4 ,clip]{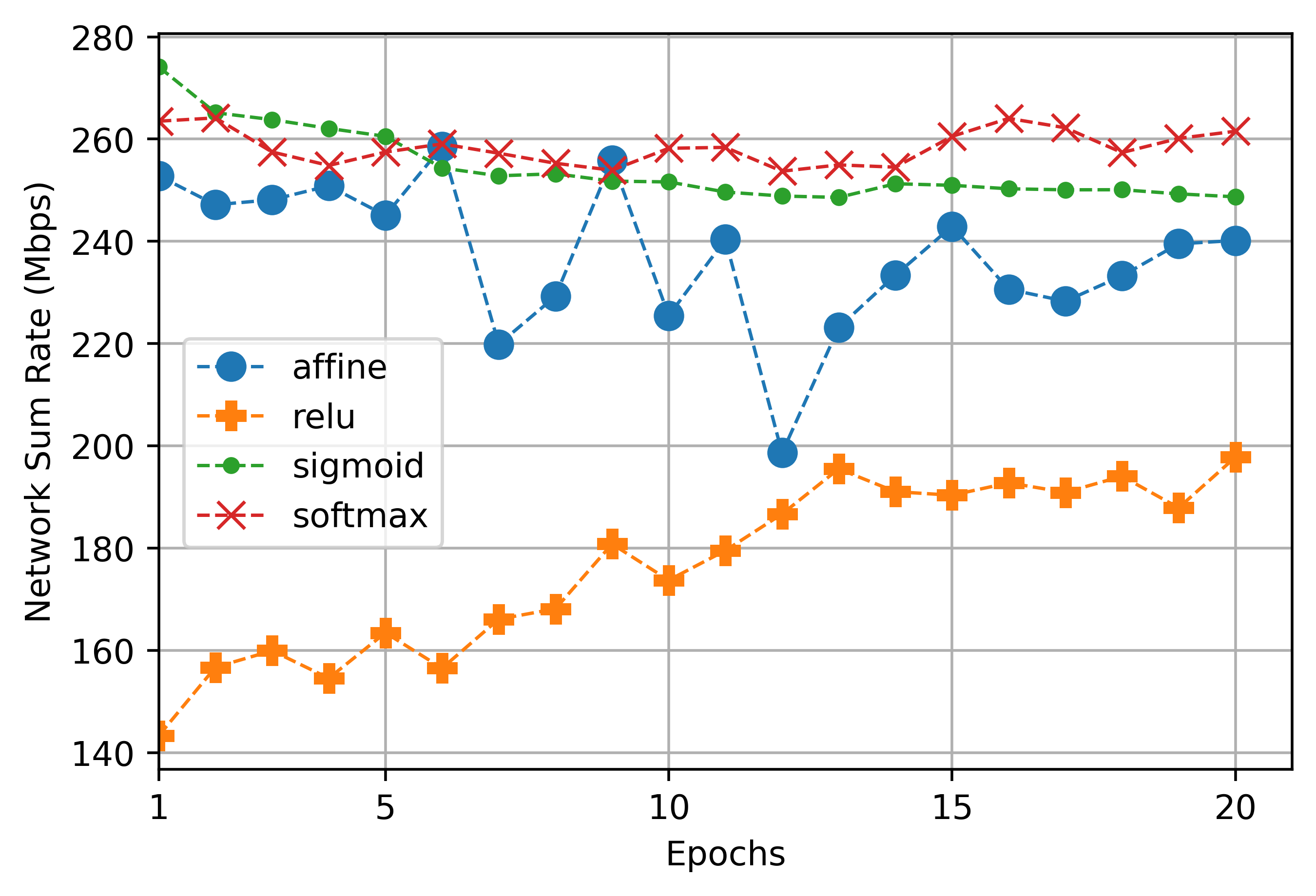}
	\end{minipage}\hfill
	\begin{minipage}{0.5\textwidth}
		\centering
		\includegraphics[scale=0.5,  trim=4 4 4 4 ,clip]{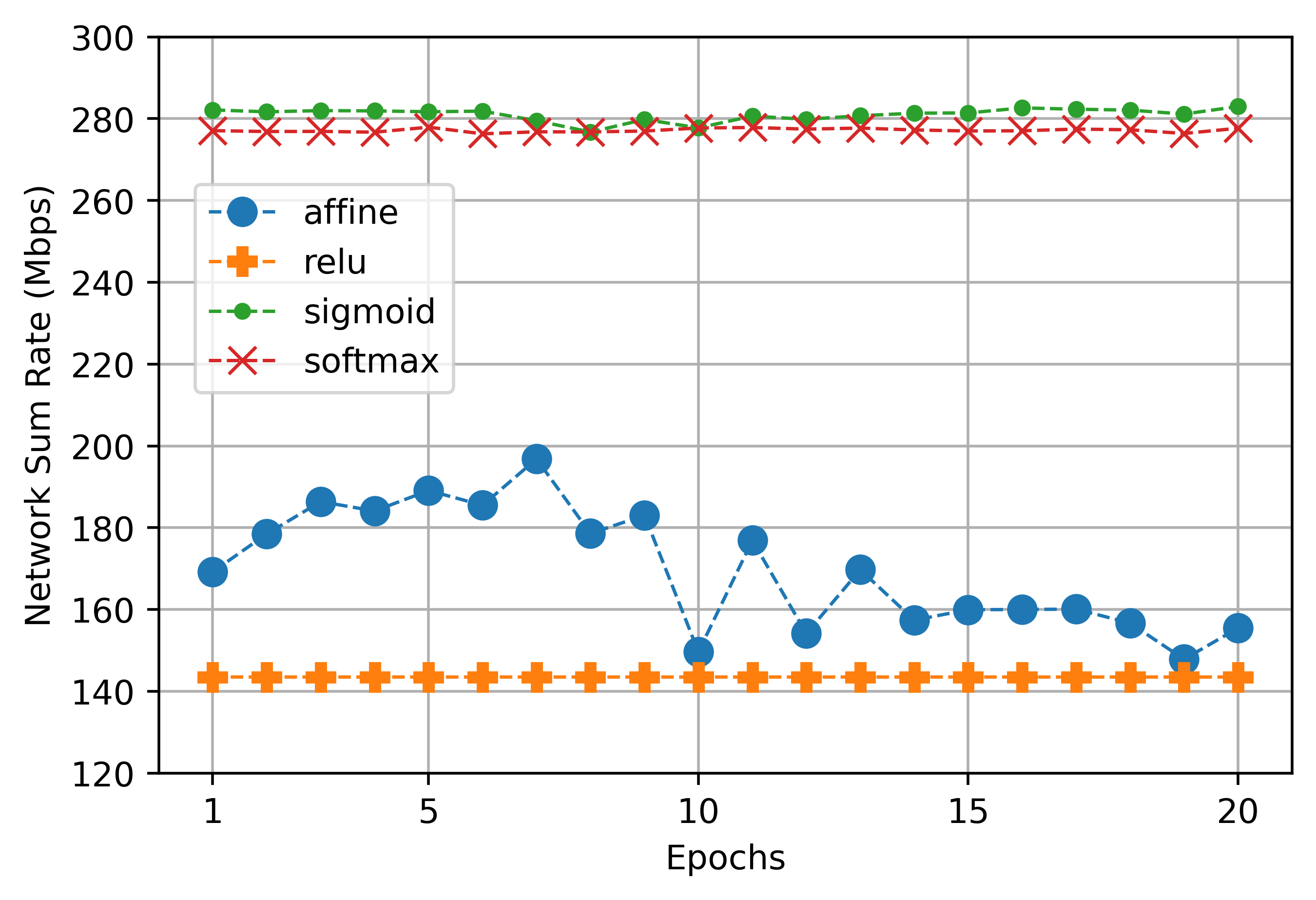}
	\end{minipage}
	\caption{The comparison of sum-rate during training using different activation functions for the last layer of (Left: DIPNet - Right: DEPNet).}
	\label{fig:out-layer:SR:dip}
\end{figure*}

When it comes to sum-rate, the performance of DIPNet is slightly better than DEPNet in Gaussian datasets but is worse for the Path-loss dataset. This is  because the iterative process used to define the projection function in DEPNet provides better gradients for the backbone neural network. The same trend can be observed in Table \ref{table:more}. Although tempting, the better performance of DEPNet comes at the cost of the careful configuration of the parameters of its projection function.

\begin{figure*}[t]
	\begin{minipage}{0.5\textwidth}
		\centering
		\includegraphics[scale=0.5,  trim=4 4 4 4 ,clip]{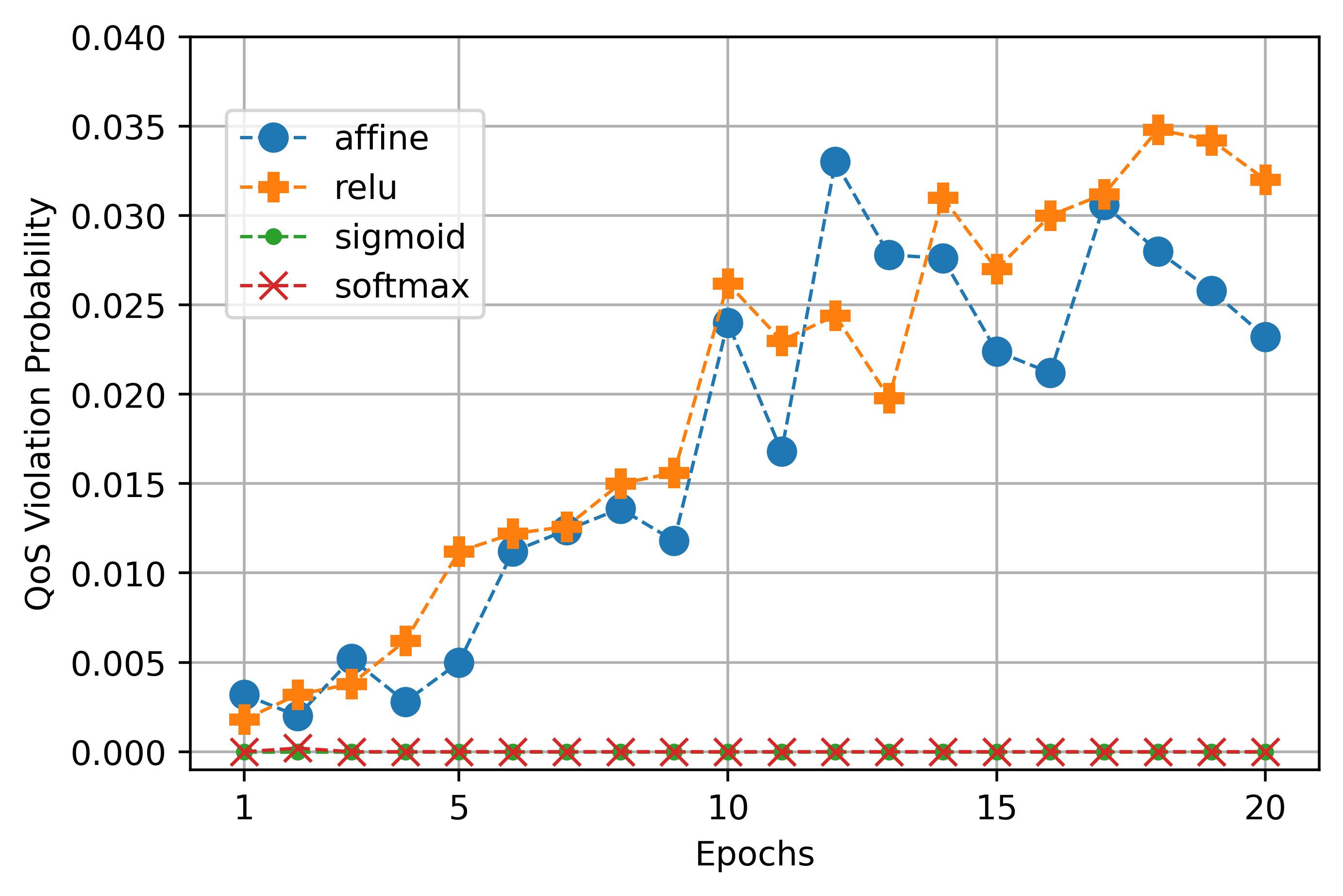}
	\end{minipage}
	\begin{minipage}{0.5\textwidth}
		\centering
		\includegraphics[scale=0.5,  trim=4 4 4 4 ,clip]{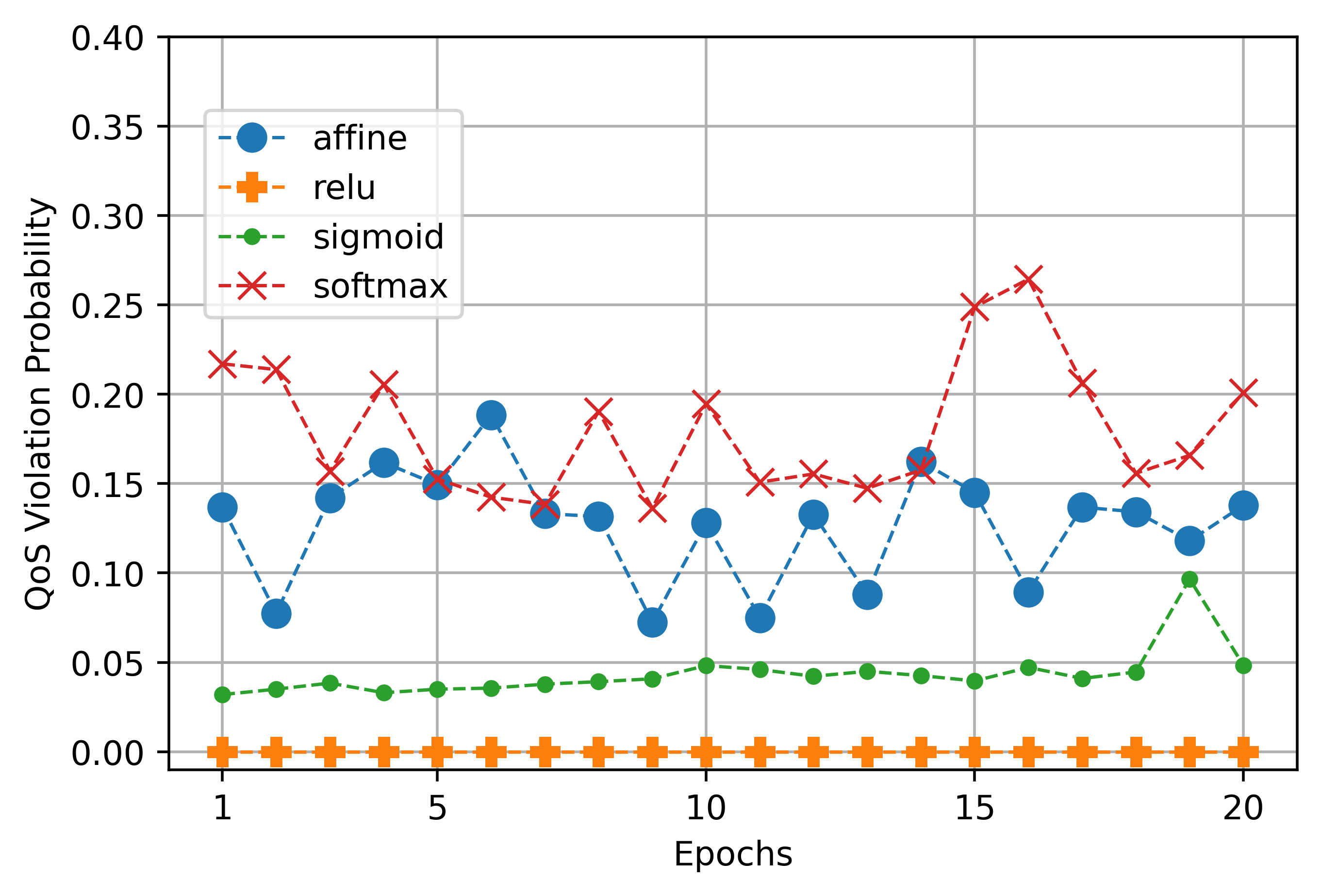}
	\end{minipage}
	\caption{The comparison of constraint violation probability during training using different activation functions for the last layer of (Left: DIPNet - Right: DEPNet).}
	\label{fig:out-layer:vio:dip}
\end{figure*}

\begin{figure*}[t]
	\begin{minipage}{0.32\textwidth}
		\includegraphics[scale=0.35,  trim=4 4 4 4 ,clip]{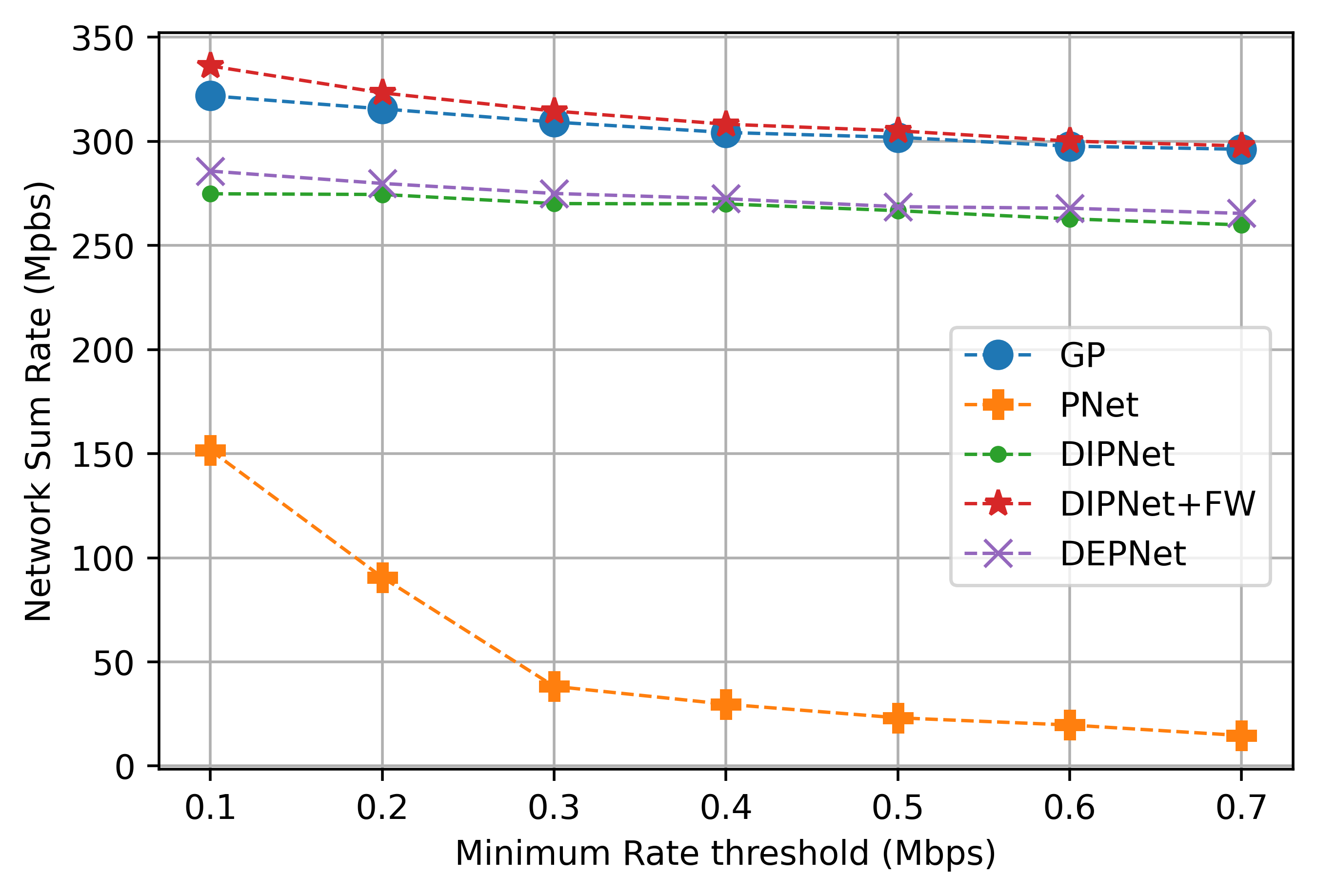}
	\end{minipage}
	\begin{minipage}{0.32\textwidth}
		\includegraphics[scale=0.35,  trim=4 4 4 4 ,clip]{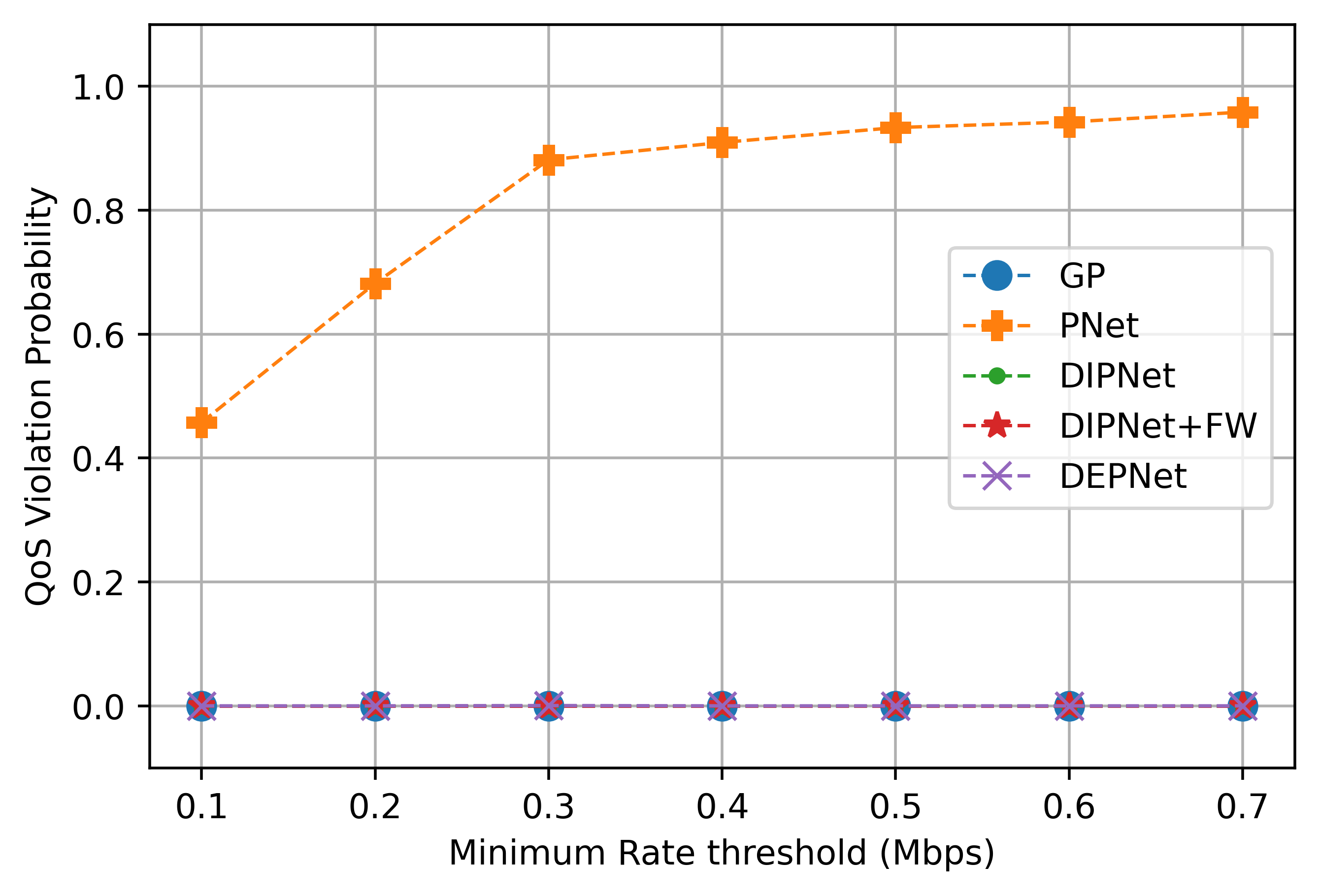}
	\end{minipage}
	\begin{minipage}{0.32\textwidth}
		\includegraphics[scale=0.35,  trim=4 4 4 4 ,clip]{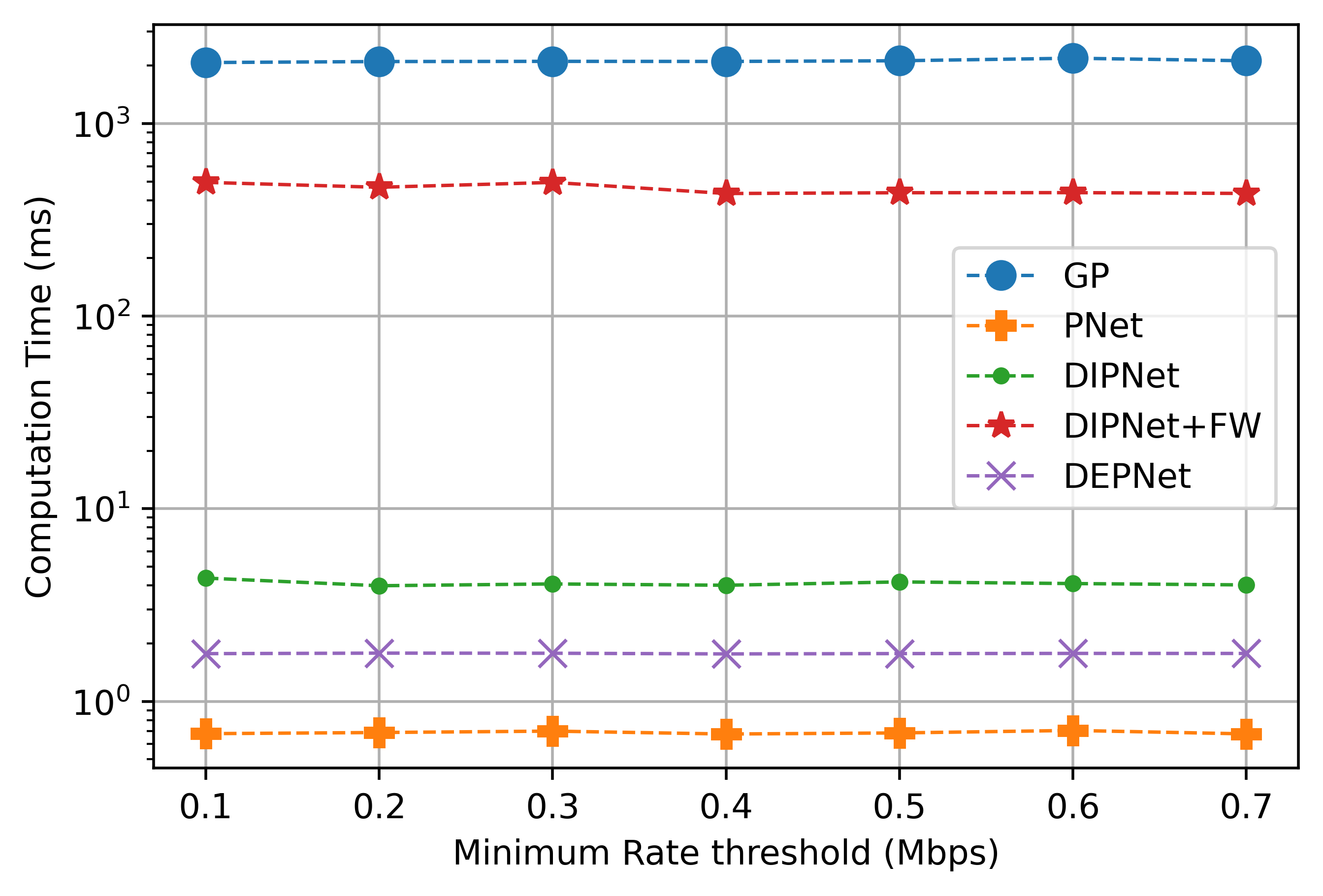}
	\end{minipage}
	\caption{Average network sum rate (left), QoS violation probability (middle), and computation time (right) as a function $\alpha_{b,q}$ considering path-loss dataset, $B=4$, $U=12$.}
	\label{fig:rate}
\end{figure*}

\begin{figure*}[t]
	\begin{minipage}{0.32\textwidth}
		\includegraphics[scale=0.35,  trim=4 4 4 4 ,clip]{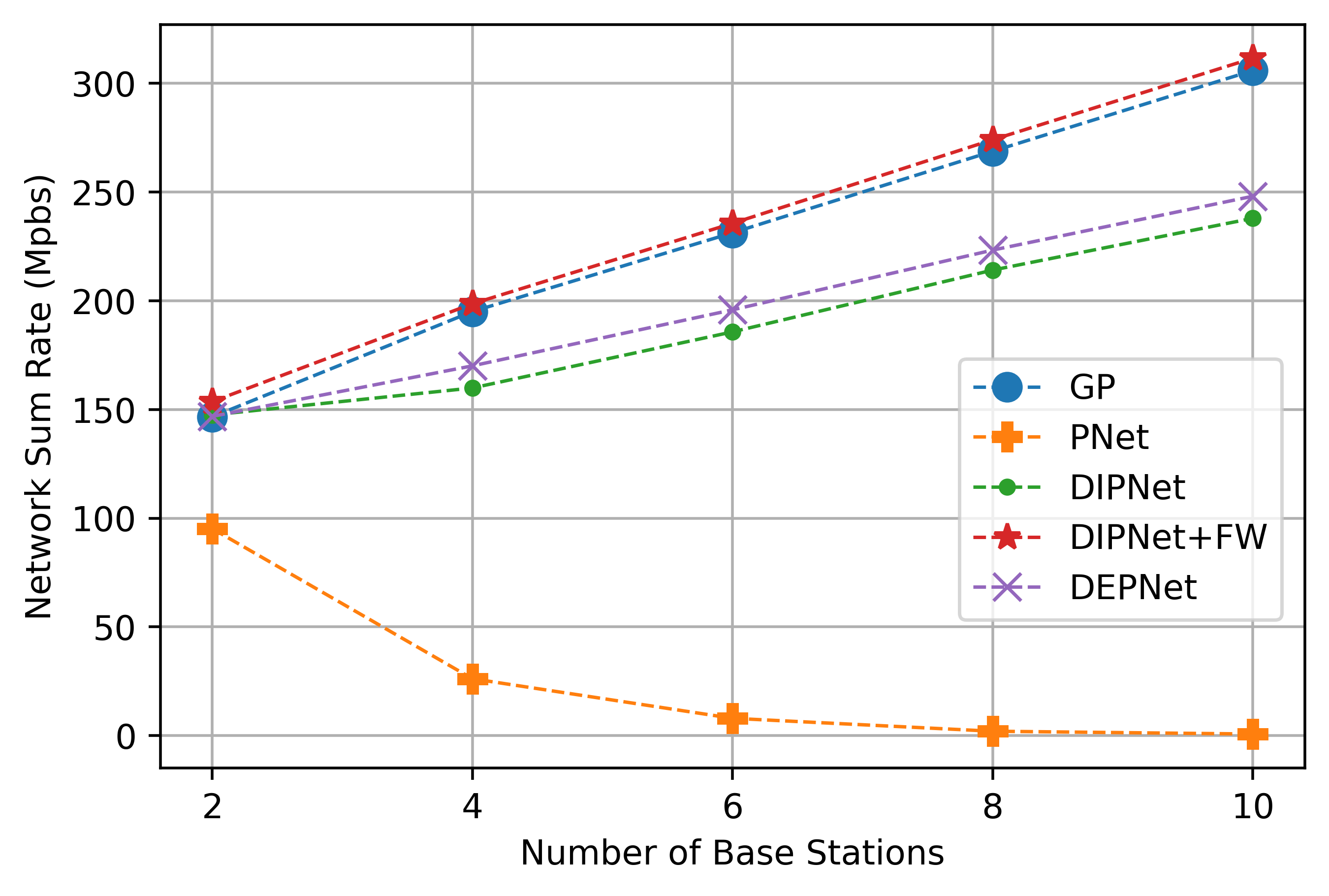}
	\end{minipage}
	\begin{minipage}{0.32\textwidth}
		\includegraphics[scale=0.35,  trim=4 4 4 4 ,clip]{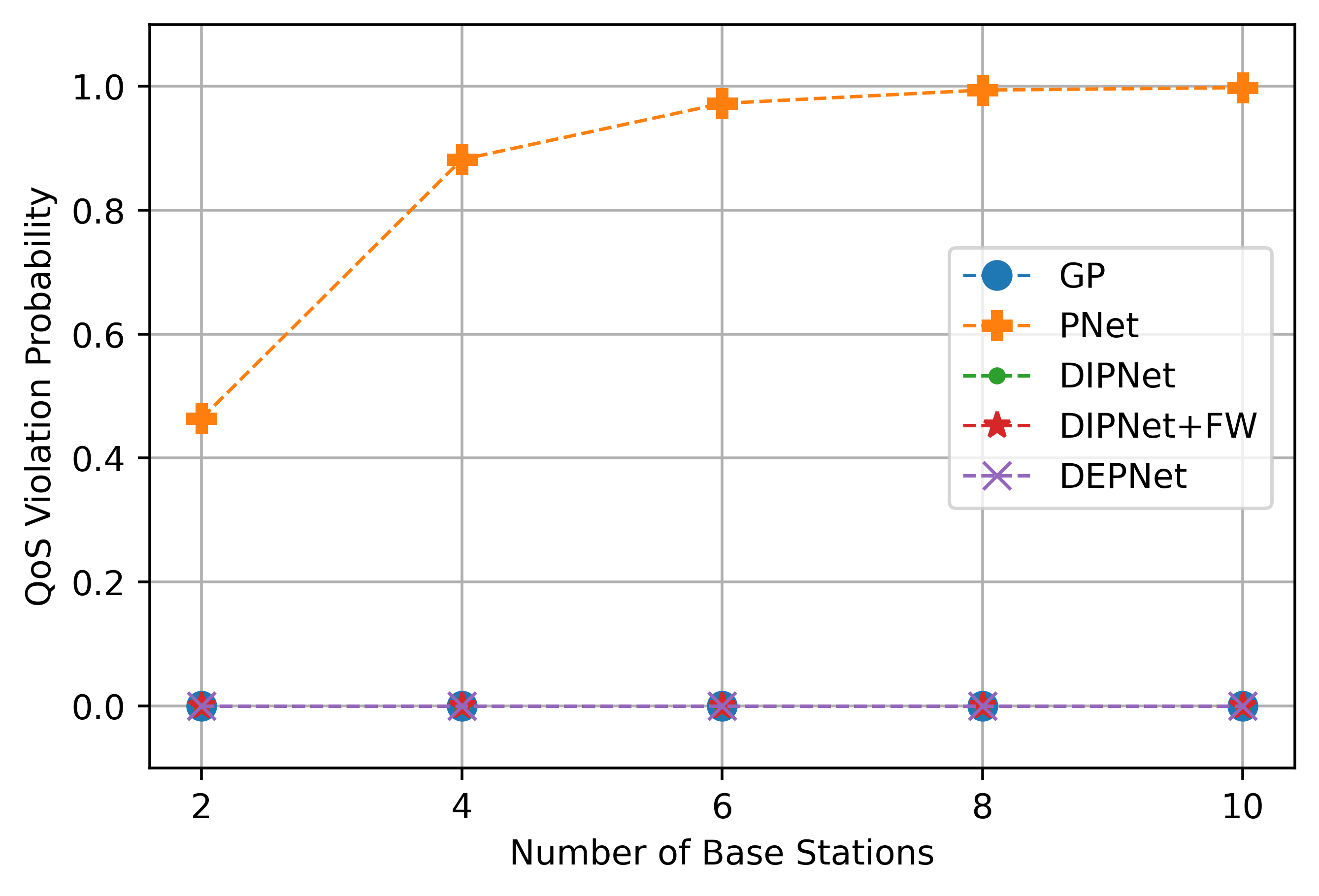}
	\end{minipage}
	\begin{minipage}{0.32\textwidth}
		\includegraphics[scale=0.35,  trim=4 4 4 4 ,clip]{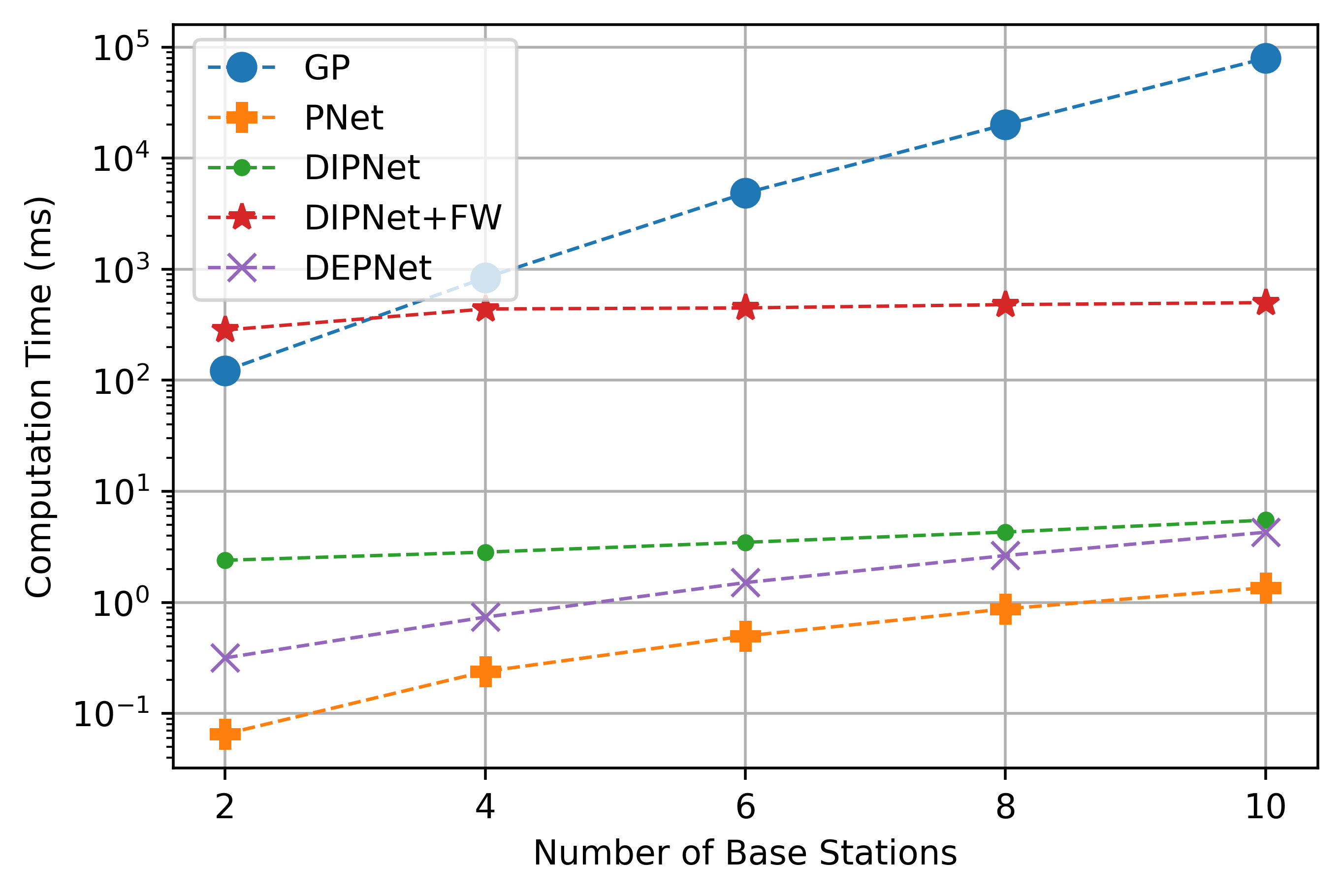}
	\end{minipage}
	\caption{Average network sum rate (left), QoS violation probability (middle), and computation time (right) as a function the $B$ considering path-loss dataset, $Q = 2$, $U=B \times Q$.}
	\label{fig:bs}
\end{figure*}

The computation time of DEPNet is faster than DIPNet in all the experiments consistently (Fig. \ref{time}). This is because DIPNet requires solving a quadratic program in each step. Moreover, since the iterative process used in DEPNet is based on Newton method and only requires gradient, Hessian, matrix inversion, and matrix multiplication, DEPNet can be run on GPU as well, which can significantly improve the computation time of DEPNet. DIPNet, however, uses an optimization solver for the projection function, which is not GPU-friendly. Hence, DEPNet outperforms DIPNet in terms of time complexity, but DIPNet is relatively easier to implement. {Fig. \ref{fig:rate} and Fig. \ref{fig:bs} show the performance of the proposed models against the benchmarks on different values of the users' minimum rate requirement and number of BSs, respectively. The general trends are observed to be the same, i.e., DIPNet and DEPNet reach zero constraint violation probability while having low computational complexity than the other benchmarks.}

\subsection{DIPNet and DEPNet: Tuning}
\subsubsection{Effect of the Activation function of the last Layer of $\mathcal{N}_r$}
Here, we examine the effect of using different activation functions for the last layer of $\mathcal{N}_r$ before the projection layer for DIPNet and DEPNet. We can think of the output of the $\mathcal{N}_r$ as the initial point for the projection layer; thus, using a proper activation function can benefit the convergence behaviour of the projection function, especially the iterative projection of DEPNet. The experiments are conducted on dataset ID 3 with $U=12$ that follows the Path-loss model. For each activation,  we record the network sum-rate and QoS violation probability of the model on the test dataset after each epoch.  We test the following activations: affine, i.e. not using any activation, ReLU, Sigmoid, and Softmax, where it is applied in a way to satisfy the second constraint of \eqref{prob:main:vec} ($\textbf{A}\textbf{r} \leq P_{\mathrm{max}}\textbf{1}$). Starting with the network sum-rate, we observe that Sigmoid and Softmax outperform other activation functions for both DIPNet and DEPNet (as shown in Fig. \ref{fig:out-layer:SR:dip}). We can see that sigmoid reaches higher network sum-rate and least QoS violation for DIPNet and DEPNet. Thus, we conclude that Sigmoid has the best performance overall.     

\subsubsection{Tuning  DEPNet: Gradient Descent vs Newton}

We compare now the convergence rate of gradient-descent (GD) with different step-sizes and Newton method in a DEPNet.  The input to the algorithms ($\textbf{r}$) is the output of the backbone neural network ($\mathcal{N}_r$) before training and after training. The training is conducted with the step-size of $0.007$ as it has the best convergence rate among the others. Moreover, the momentum is set to 0.5 for all the gradient-descent configurations, and $10^{-8}$ is used to regularize the Hessian. 

\begin{figure*}[t]
	\centering
	\begin{minipage}{0.5\textwidth}
		\centering
		\includegraphics[scale=0.55,  trim=4 4 4 4 ,clip]{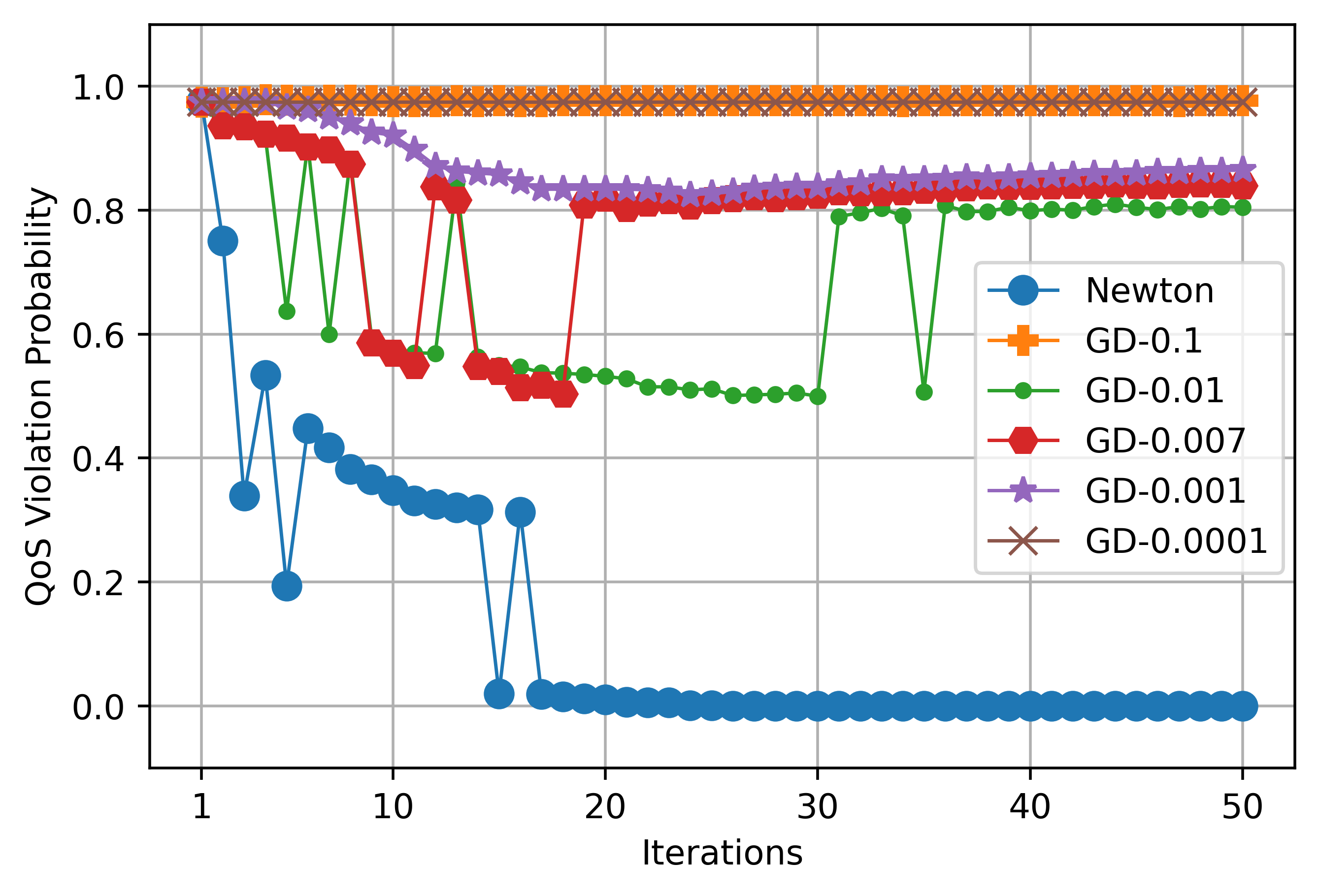}
		\label{fig:converg:before}
	\end{minipage}\hfill
	\begin{minipage}{0.5\textwidth}
		\centering
		\includegraphics[scale=0.55,  trim=4 4 4 4 ,clip]{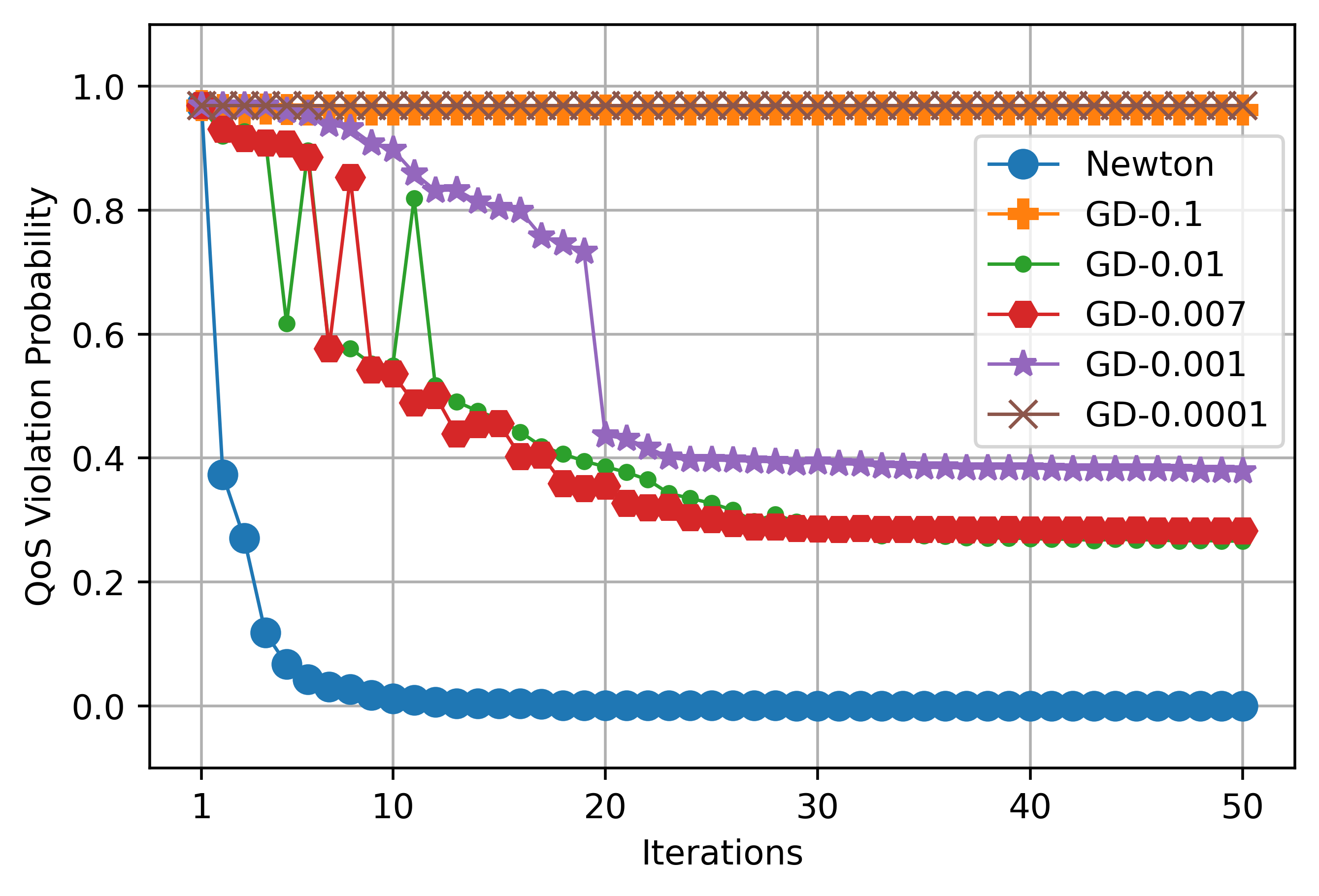}
		\label{fig:converg:after}
	\end{minipage}
	\caption{Convergence of DEPNet considering GD with different step sizes and Newton method (Left: before training- Right: after training).}
	\label{fig:converg}
\end{figure*}


Fig. \ref{fig:converg} shows 
\textcolor{black}{the dynamics of the projection function before and after training considering Newton and GD (with different step sizes). It is observed that all configurations exhibit an improvement in convergence after training, thus indicating that the backbone neural network learns to generate an initial point for the projection function that is already in proximity to the feasible set.  However, the fluctuations in Fig. 11 (especially in gradient descent method after training) are attributed to the fixed step size and momentum employed in each iteration of the gradient descent. The rationale behind this choice is to ensure computational efficiency and ease of  differentiability for the training of the backbone neural network through back-propagation. To minimize fluctuations, one solution is to adopt an adaptive step-size by performing a line search \cite{boyd2004convex}. However, this approach increases the computational complexity of each  gradient descent update and complicates differentiation, rendering its application challenging during training.}

As we can see, Newton method is the only correction process that can achieve zero violation probability. Among different step-sizes for GD, we can observe that step-size 0.01 and 0.007 achieve lower violation probability than others. We can see that having very large and small step-sizes (0.1 and 0.0001) results in no progress. Thus, the step-size should be chosen with experiments to find the right range that helps the convergence of the projection function. 



\begin{figure*}[t]
	\begin{minipage}{0.5\textwidth}
		\centering
		\includegraphics[scale=0.5,  trim=4 4 4 4 ,clip]{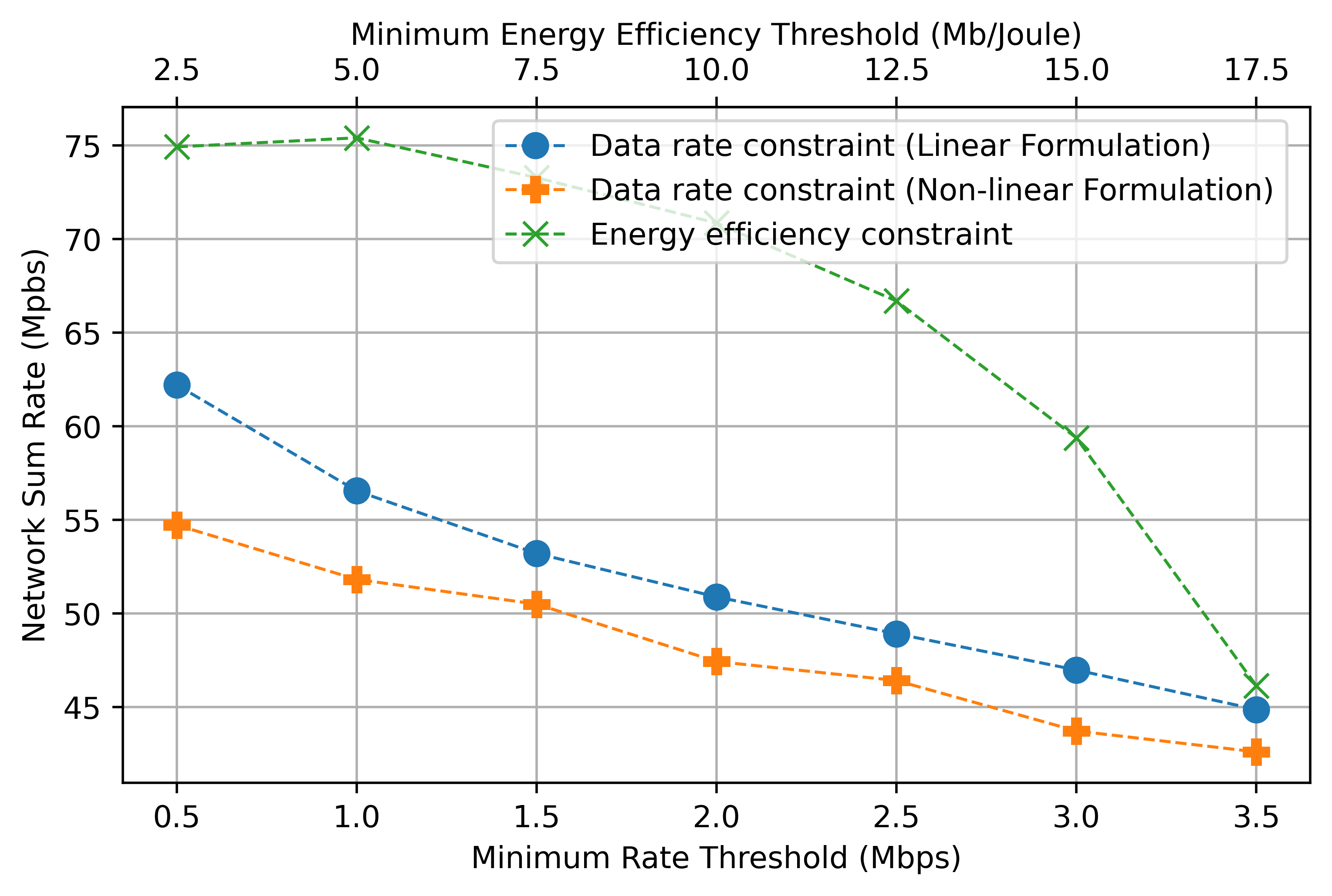}
	\end{minipage}\hfill
	\begin{minipage}{0.5\textwidth}
		\centering
		\includegraphics[scale=0.5,  trim=4 4 4 4 ,clip]{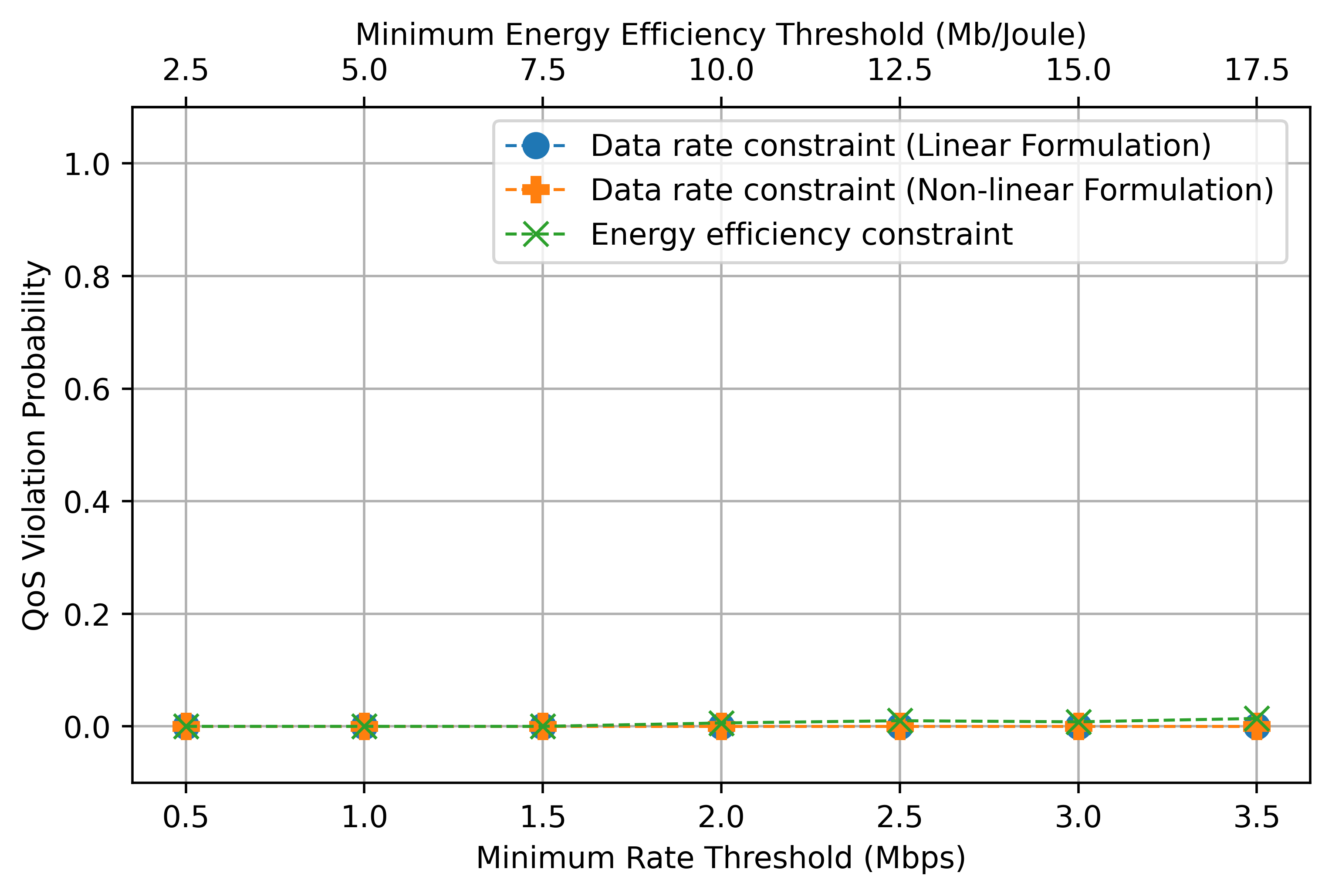}
	\end{minipage}
	\caption{Average network sum-rate (left) and QoS violation (right) for DEPNet with convex and non-convex constraints, $B=4$, $U=8$.}
	\label{fig:nonlin}
\end{figure*}

After the training is completed, we can see that the gradient-descent with step-size 0.01, 0.007, and 0.001, achieves better results than before (without training).
The Newton method still is the winner of the game by achieving zero violation probability. We can see that the convergence of Newton method is improved after training and it converges after almost 10 iterations, which took about 20 iterations to happen before the training. Moreover,  after training, we don't have any oscillations before the convergence. This implies that by training, the backbone network learns to output points that are already very close to the feasible set, and just taking a few steps of Newton method will land them on the feasible set.
\textcolor{black}{
	\subsection{Consideration of Non-Convex Constraints}
	We would like to emphasize that the DIPNet requires convex constraints; whereas DEPNet does not pose any such restriction. That is, DEPNet can handle more sophisticated non-convex constraints such as energy efficiency constraints. To demonstrate this capability, we tested DEPNet on three network sum rate maximization problems: 1)  with non-linear data rate constraint formulation,  2)  with (matrix-based) linear data rate constraint formulation, and 3)  with energy efficiency (EE) constraint. Since the objective function and power budget constraint are the same among these problems, in the following, we only show the QoS or EE constraint:
	\textbf{(i)}~$\text{Data Rate (non-linear):} R_{b,q}(\textbf{P},\textbf{H})~\geq~\alpha_{b,q},\forall b\in\mathcal{B}, \forall q\in\mathcal{Q},\:
	\textbf{(ii)}~\text{Data Rate (linear):}~\textbf{C}\textbf{p}~\geq~\textbf{d}, \:
	\textbf{(iii)}EE_{b,q} = \frac{R_{b,q}(\textbf{P}, \textbf{H})}{P_{b,q}} \geq ee_{b,q}, \quad \forall b \in \mathcal{B}, \forall q \in \mathcal{Q}$,
	where $ee_{b,q}$ is the minimum energy efficiency required by the user  associated with channel $q$ of BS $b$. The general procedure of using DEPNet remains the same as in Section~V. 
	As we can see in Fig. \ref{fig:nonlin} (right), DEPNet works successfully for all of the cases and reaches zero violation for data rate constraint with both linear and non-linear constraints. The number of iterations of the newton method is set to 100 for the linear one and 300 for the other two. The reason behind this difference is that the other two constraints involve non-convexity, which requires the newton method to take more steps to satisfy them. Finally, as expected, the data rate and energy efficiency continue to decrease with the increase in data rate requirements.  The reason is that the gains from power allocations to strong channel users continue to diminish as the power demands of weak channel users continue to increase; thus overall data rate decreases.
}

\begin{figure*}[t]
	\begin{minipage}{0.5\textwidth}
		\centering
		\includegraphics[scale=0.5,  trim=4 4 4 4 ,clip]{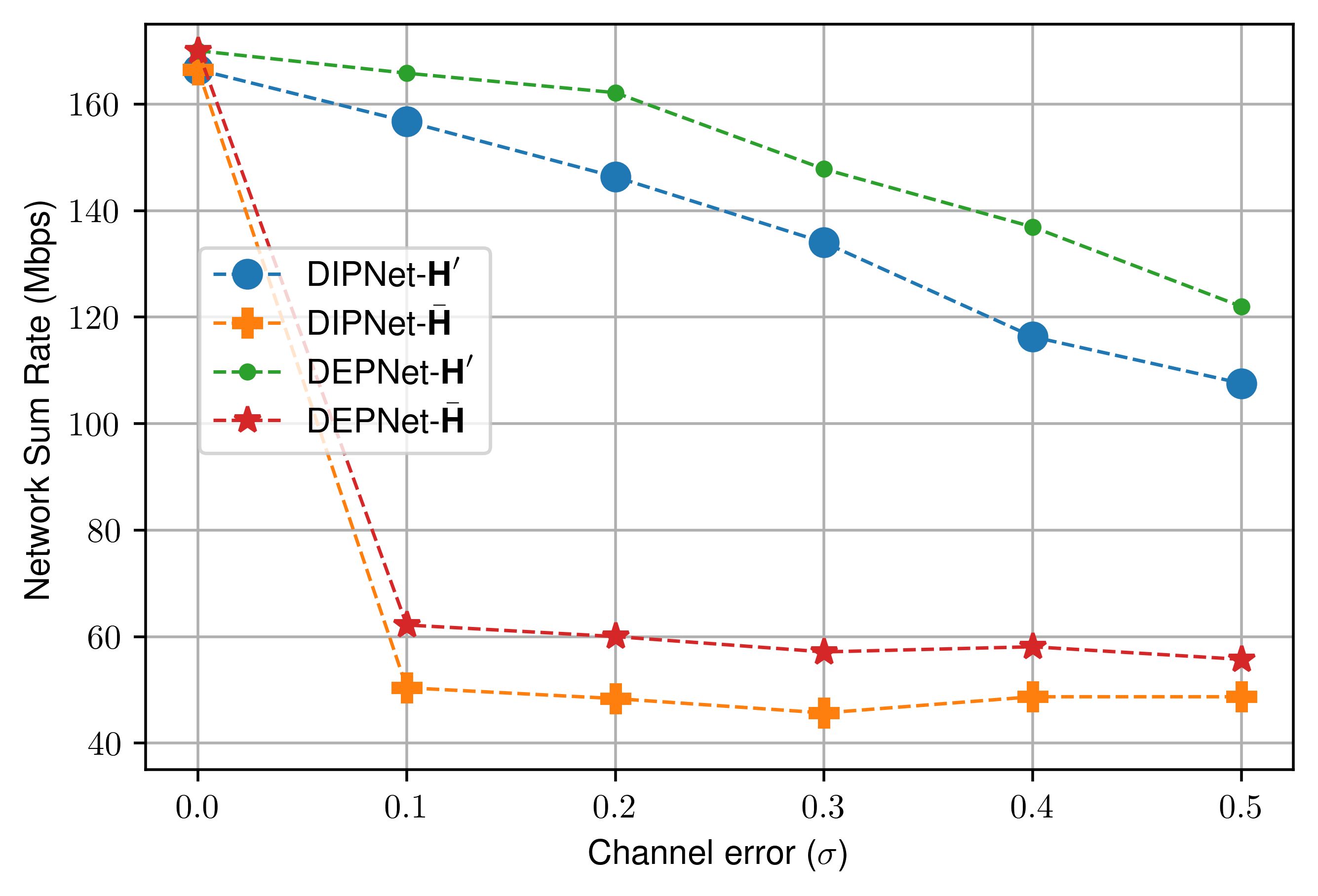}
	\end{minipage}
	\begin{minipage}{0.5\textwidth}
		\centering
		\includegraphics[scale=0.5,  trim=4 4 4 4 ,clip]{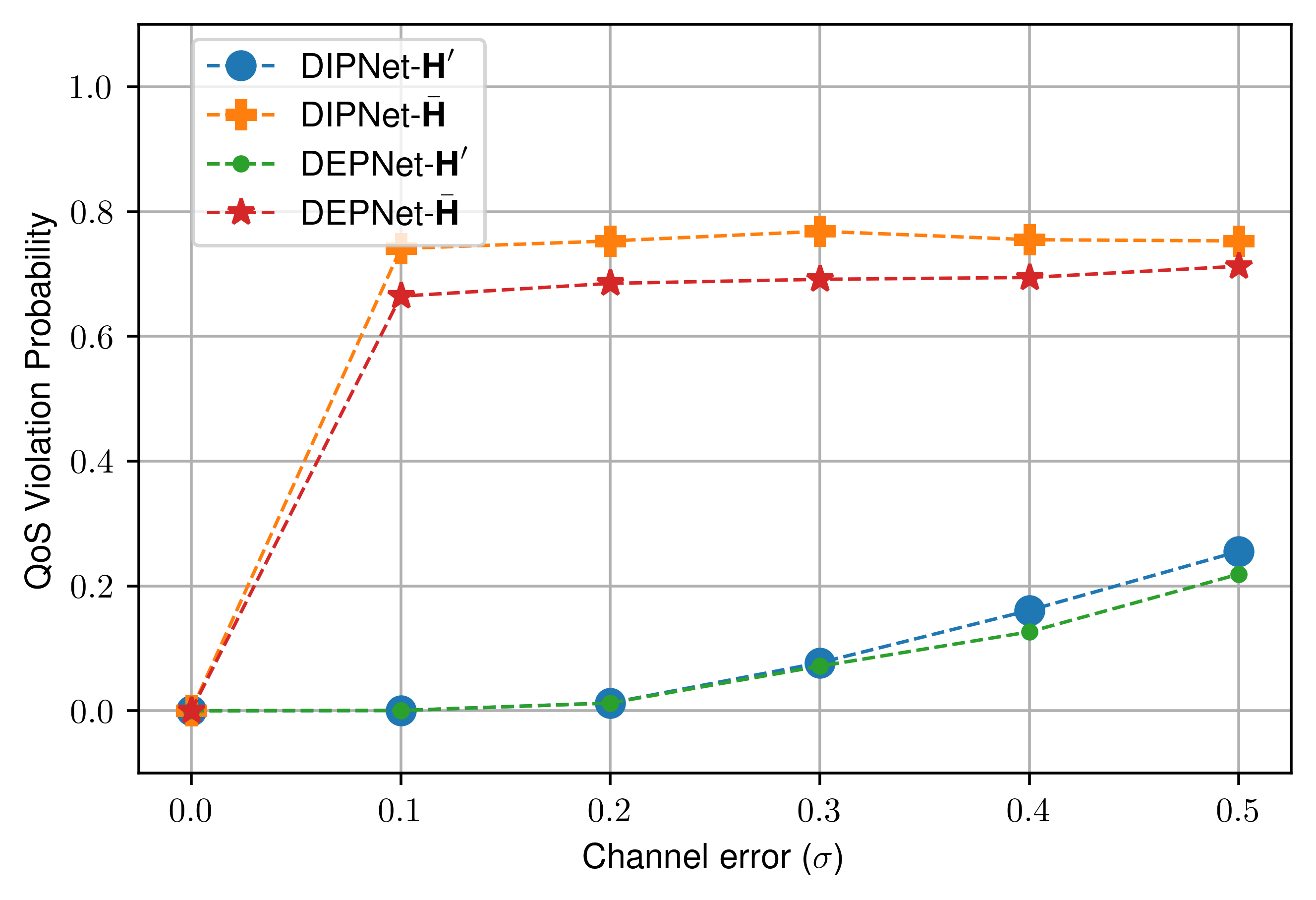}
	\end{minipage}
	\caption{Average network sum-rate (left) and QoS violation probability (right) for DIPNet and DEPNet in the presence of imperfect CSI considering path-loss dataset, $B$=4, $U$=8, $\alpha_{b,q} =2.5$ Mbps.}
	\label{fig:imperfect}
\end{figure*}
\subsection{Consideration of Imperfect CSI}
Following the CSI estimation error model in \cite{lee2021robust}, we have:
$
H_{b,q,\hat{b}} = \hat{H}_{b,q,\hat{b}} + \Delta H_{b,q,\hat{b}}, \quad \mathrm{   where} \quad\Delta H_{b,q,\hat{b}} \sim \mathcal{U}[-\sigma H_{b,q,\hat{b}}, \sigma H_{b,q,\hat{b}}]
$
where $\hat{H}_{b,q,\hat{b}}$ is the estimated imperfect CSI that the DNN has access to (both in training and test time) and $\Delta H_{b,q,\hat{b}}$ is the estimation error following a uniform distribution. The uncertainty increases with the increase in $\sigma$, which widens the support of the uniform distribution. To overcome the estimation error in the CSI, \cite{lee2021robust} proposed to generate an artificially distorted CSI  which has the same statistical characteristics as $\Delta H_{b,q,\hat{b}}$, and use it as the input to the loss function of the DNN. In other words,
$
\bar{H}_{b,q,\hat{b}} = \hat{H}_{b,q,\hat{b}} + \Delta \hat{H}_{b,q,\hat{b}},  \mathrm{   where} \quad \Delta \hat{H}_{b,q,\hat{b}} \sim \mathcal{U}[-\sigma \hat{H}_{b,q,\hat{b}}, \sigma \hat{H}_{b,q,\hat{b}}]
$
where we denote the tensors of imperfect, artificially distorted, and perfect CSIs, as $\hat{\textbf{H}}$, $\bar{\textbf{H}}$, and $\textbf{H}$, respectively. As shown in \cite{lee2021robust}, using the artificially distorted CSI during the training will make the output of the DNN robust to the estimation error. In Fig.~\ref{fig:imperfect}, we note that the quality of the CSI influences the amount of the resulting sum-rate. Moreover, as the estimation error increases, the QoS violation  increases.  The reason is that the proposed projection methods are designed to find a point in the feasible set to fulfill the constraints with zero violation, and the geometry of the feasible set is a function of the input CSI to the DNN, thus considering imperfect CSI results in a constraint violation.

\textcolor{black}{To make the proposed projection methods behave in a robust manner w.r.t. estimation error in CSI, we propose  a heuristic to tackle the CSI imperfection (as detailed in the appendix). }
To test the approach, we generate $\bar{\textbf{H}}$ and $\hat{{\textbf{H}}}$ for each value of $\sigma$ and applied the feasibility check to make sure all the data points are feasible.  The hyperparameters for training DIPNet and DEPNet are the same as the other results of the paper. To find the worst-case, we generate multiple CSIs with $\chi$ ranging from zero to $\sigma$ with a step size of 0.01, apply the feasibility check on all of them and pick the one with the largest value of $\chi$. This will provide us with an approximation of $\textbf{H}_{\mathrm{min}}$. As shown in Fig. \ref{fig:imperfect}, this approach works well for small values of $\sigma$ for both DIPNet and DEPNet and results in zero constraint violation probability. As $\sigma$ increases, however, we observe some violations, which are still significantly lower than the method in \cite{lee2021robust}. The reason is  using the same value of $\chi$ for all the CSIs to approximate the worst-case scenario, which is an approximation of  $\textbf{H}_{\mathrm{min}}$. One can find the exact value of $\textbf{H}_{\mathrm{min}}$ by following sophisticated techniques from robust optimization in \cite{ben2009robust}.

\section{Conclusion}
In this paper, to achieve zero constraint violation probability, a differentiable projection framework is developed, which uses a projection function to project the output of the backbone neural network to the feasible set of the problem. The projection function is defined implicitly using convex optimization and explicitly using an iterative process. The resulting DIPNet and DEPNet are tested against optimization-based and neural-based benchmarks. Numerical experiments confirmed zero violation probability of the output of the proposed models while outperforming the DNN-based benchmark in terms of sum-rate and GP in terms of the computation time. 
\textcolor{black}{With the proposed framework, one can handle more sophisticated differentiable constraints that are a function of problem data and/or the constraints that cannot be handled with the standard off-the-shelf projection functions.  To incorporate non-differentiable constraints in our framework, one can apply some continuous approximations to provide differentiability. This is synonymous with the work in \cite{liu2022deep}, where the categorical distribution is approximated with Gumble-Softmax distribution, and the quantization function is approximated during training with a smooth function to make the constraints differentiable; thereby, the training of the DNN becomes possible. Another way to extend the proposed framework for non-differentiable constraints is to design another differentiable and iterative process specific to those constraints, to satisfy them. One example of such work is \cite{kim2021deep}, where the authors used an iterative process called Sinkhorn normalization to  project the output of the neural network to the space of doubly-stochastic matrices, i.e. positive-valued square matrices where the sum of each row and column is one.} \textcolor{black}{Moreover, considering imperfect CSI is inevitable due to the imperfect channel estimation procedures. Thus, this is an important problem for further investigation. \textcolor{black}{Furthermore, there are several other important QoS performance metrics like energy efficiency, fairness, reliability, latency, jitter, that can be investigated with the proposed framework. }}

\begin{appendix}
	Start from the linear formulation of the constraints in \eqref{prob:main:vec}. 
	As we can see, the matrix $\textbf{C}$ is derived directly from the CSIs (we denote it as $\textbf{C}(\textbf{H})$). Thus, if we design the projection w.r.t. $\hat{\textbf{H}}$, there is a certain probability that it will not work for $\textbf{H}$. 
	To overcome this issue, given the estimated CSI and the distribution of the estimation error, we can generate the worst-case CSI  (denoted by $\textbf{H}_{\mathrm{min}}$) that is feasible w.r.t. the minimum rate requirement of the users.  Thus, once we perform the projection w.r.t.  $\textbf{H}_{\mathrm{min}}$, the constraints will be satisfied for the perfect CSI as well. In other words, given $\hat{\textbf{H}}$, we want to find $\textbf{H}_{\mathrm{min}}$ such that if
	$
	\textbf{C}(\textbf{H}_{\mathrm{min}}) \textbf{p} \geq \textbf{d} \Rightarrow \textbf{C}(\textbf{H}) \textbf{p} \geq \textbf{d}.
	$
	In the following, we generate an approximation of $\textbf{H}_{\mathrm{min}}$, i.e., ($\textbf{H}'$) by making the direct channel of each user weaker and the interfering channels stronger. Mathematically:
	\begin{equation}
		\centering
		\begin{aligned}
			H'_{b,q,\hat{b}} = 
			\left\{
			\begin{array}{ll}
				\hat{H}_{b,q,b} - \chi \hat{H}_{b,q,b} & \mbox{if } \hat{b} = b\\
				\hat{H}_{b,q,b} + \chi \hat{H}_{b,q,b} & \mbox{otherwise}
			\end{array} \right.
		\end{aligned}    
	\end{equation}
	where $0 < \chi \leq \sigma$.
\end{appendix}
\bibliographystyle{IEEEtran}
\bibliography{ref.bib}

\begin{thebibliography}{10}
\providecommand{\url}[1]{#1}
\csname url@samestyle\endcsname
\providecommand{\newblock}{\relax}
\providecommand{\bibinfo}[2]{#2}
\providecommand{\BIBentrySTDinterwordspacing}{\spaceskip=0pt\relax}
\providecommand{\BIBentryALTinterwordstretchfactor}{4}
\providecommand{\BIBentryALTinterwordspacing}{\spaceskip=\fontdimen2\font plus
\BIBentryALTinterwordstretchfactor\fontdimen3\font minus
  \fontdimen4\font\relax}
\providecommand{\BIBforeignlanguage}[2]{{%
\expandafter\ifx\csname l@#1\endcsname\relax
\typeout{** WARNING: IEEEtran.bst: No hyphenation pattern has been}%
\typeout{** loaded for the language `#1'. Using the pattern for}%
\typeout{** the default language instead.}%
\else
\language=\csname l@#1\endcsname
\fi
#2}}
\providecommand{\BIBdecl}{\relax}
\BIBdecl

\bibitem{liu2012achieving}
L.~Liu, R.~Zhang, and K.-C. Chua, ``Achieving global optimality for weighted
  sum-rate maximization in the {K}-user {G}aussian interference channel with
  multiple antennas,'' \emph{IEEE Trans. on Wireless Commun.}, vol.~11, no.~5,
  pp. 1933--1945, 2012.

\bibitem{shi2011iteratively}
Q.~Shi, M.~Razaviyayn, Z.-Q. Luo, and C.~He, ``An iteratively weighted {MMSE}
  approach to distributed sum-utility maximization for a {MIMO} interfering
  broadcast channel,'' \emph{IEEE Trans. on Signal Processing}, vol.~59, no.~9,
  pp. 4331--4340, 2011.

\bibitem{liang2019towards}
F.~Liang, C.~Shen, W.~Yu, and F.~Wu, ``Towards optimal power control via
  ensembling deep neural networks,'' \emph{IEEE Trans. on Commun.}, vol.~68,
  no.~3, pp. 1760--1776, 2019.

\bibitem{kaushik2021deep}
A.~Kaushik, M.~Alizadeh, O.~Waqar, and H.~Tabassum, ``Deep unsupervised
  learning for generalized assignment problems: A case-study of
  user-association in wireless networks,'' in \emph{IEEE Intl. Conf. on Commun.
  Wkshps. (ICC Wkshps.)}, 2021.

\bibitem{added1}
K.~Lee, J.-P. Hong, H.~Seo, and W.~Choi, ``Learning-based resource management
  in device-to-device communications with energy harvesting requirements,''
  \emph{IEEE Trans. on Commun.}, vol.~68, no.~1, pp. 402--413, 2020.

\bibitem{sun2019learning}
C.~Sun and C.~Yang, ``Learning to optimize with unsupervised learning: Training
  deep neural networks for urllc,'' in \emph{IEEE 30th Annual Intl. Symposium
  on Personal, Indoor and Mobile Radio Commun. (PIMRC)}, 2019, pp. 1--7.

\bibitem{added}
M.~Eisen, C.~Zhang, L.~F.~O. Chamon, D.~D. Lee, and A.~Ribeiro, ``Learning
  optimal resource allocations in wireless systems,'' \emph{IEEE Trans. on
  Signal Processing}, vol.~67, no.~10, pp. 2775--2790, 2019.

\bibitem{sun2018learning}
H.~Sun, X.~Chen, Q.~Shi, M.~Hong, X.~Fu, and N.~D. Sidiropoulos, ``Learning to
  optimize: Training deep neural networks for interference management,''
  \emph{IEEE Trans. on Signal Processing}, vol.~66, no.~20, pp. 5438--5453,
  2018.

\bibitem{deng2019application}
Z.~Deng, Q.~Sang, Y.~Pan, and Y.~Xin, ``Application of deep learning for power
  control in the interference channel: a {RNN}-based approach,'' in \emph{Conf.
  on Research in Adaptive and Convergent Systems}, 2019, pp. 96--100.

\bibitem{lee2018deep}
W.~Lee, M.~Kim, and D.-H. Cho, ``Deep power control: Transmit power control
  scheme based on convolutional neural network,'' \emph{IEEE Commun. Letters},
  vol.~22, no.~6, pp. 1276--1279, 2018.

\bibitem{shen2020graph}
Y.~Shen, Y.~Shi, J.~Zhang, and K.~B. Letaief, ``Graph neural networks for
  scalable radio resource management: Architecture design and theoretical
  analysis,'' \emph{IEEE Journal on Sel. Areas in Commun.}, vol.~39, no.~1, pp.
  101--115, 2020.

\bibitem{eisen2020optimal}
M.~Eisen and A.~Ribeiro, ``Optimal wireless resource allocation with random
  edge graph neural networks,'' \emph{ieee Trans. on signal processing},
  vol.~68, pp. 2977--2991, 2020.

\bibitem{naderializadeh2020wireless}
N.~Naderializadeh, M.~Eisen, and A.~Ribeiro, ``Wireless power control via
  counterfactual optimization of graph neural networks,'' \emph{arXiv preprint
  arXiv:2002.07631}, 2020.

\bibitem{li2021multicell}
Y.~Li, S.~Han, and C.~Yang, ``Multicell power control under rate constraints
  with deep learning,'' \emph{IEEE Trans. on Wireless Commun.}, vol.~20,
  no.~12, pp. 7813--7825, 2021.

\bibitem{9281322}
W.~Lee and K.~Lee, ``Resource allocation scheme for guarantee of qos in d2d
  communications using deep neural network,'' \emph{IEEE Commun. Letters},
  vol.~25, no.~3, pp. 887--891, 2021.

\bibitem{she2021tutorial}
C.~She, C.~Sun, Z.~Gu, Y.~Li, C.~Yang, H.~V. Poor, and B.~Vucetic, ``{A}
  {T}utorial on {U}ltrareliable and {L}ow-{L}atency {C}ommunications in {6G}:
  {I}ntegrating {D}omain {K}nowledge {I}nto {D}eep {L}earning,''
  \emph{Proceedings of the IEEE}, vol. 109, no.~3, pp. 204--246, 2021.

\bibitem{eisen2019dual}
M.~Eisen, C.~Zhang, L.~F. Chamon, D.~D. Lee, and A.~Ribeiro, ``Dual domain
  learning of optimal resource allocations in wireless systems,'' in \emph{IEEE
  Intl. Conf. on Acoustics, Speech and Signal Processing (ICASSP)}.\hskip 1em
  plus 0.5em minus 0.4em\relax IEEE, 2019, pp. 4729--4733.

\bibitem{amos2017optnet}
B.~Amos and J.~Z. Kolter, ``Optnet: Differentiable optimization as a layer in
  neural networks,'' in \emph{Intl. Conf. on Machine Learning}.\hskip 1em plus
  0.5em minus 0.4em\relax PMLR, 2017, pp. 136--145.

\bibitem{agrawal2019differentiable}
A.~Agrawal, B.~Amos, S.~Barratt, S.~Boyd, S.~Diamond, and J.~Z. Kolter,
  ``Differentiable convex optimization layers,'' \emph{Advances in neural
  information processing systems}, vol.~32, 2019.

\bibitem{toturial_imp}
Z.~Kolter, D.~Duvenaud, and M.~Johnson, ``Deep implicit layers,''
  \url{http://implicit-layers-tutorial.org}.

\bibitem{donti2021dc3}
\BIBentryALTinterwordspacing
P.~L. Donti, D.~Rolnick, and J.~Z. Kolter, ``{DC}3: A learning method for
  optimization with hard constraints,'' in \emph{Intl. Conf. on Learning
  Representations}, 2021. [Online]. Available:
  \url{https://openreview.net/forum?id=V1ZHVxJ6dSS}
\BIBentrySTDinterwordspacing

\bibitem{lacoste2016convergence}
S.~Lacoste-Julien, ``Convergence rate of frank-wolfe for non-convex
  objectives,'' \emph{arXiv preprint arXiv:1607.00345}, 2016.

\bibitem{chiang2008power}
M.~Chiang, P.~Hande, and T.~Lan, \emph{Power control in wireless cellular
  networks}.\hskip 1em plus 0.5em minus 0.4em\relax Now Publishers Inc, 2008.

\bibitem{sun2020unsupervised}
C.~Sun, C.~She, and C.~Yang, ``Unsupervised deep learning for optimizing
  wireless systems with instantaneous and statistic constraints,'' \emph{arXiv
  preprint arXiv:2006.01641}, 2020.

\bibitem{LESHNO1993861}
M.~Leshno, V.~Y. Lin, A.~Pinkus, and S.~Schocken, ``Multilayer feedforward
  networks with a nonpolynomial activation function can approximate any
  function,'' \emph{Neural Networks}, vol.~6, no.~6, pp. 861--867, 1993.

\bibitem{lecun2015deep}
Y.~LeCun, Y.~Bengio, and G.~Hinton, ``Deep learning,'' \emph{nature}, vol. 521,
  no. 7553, pp. 436--444, 2015.

\bibitem{gould2019deep}
S.~Gould, R.~Hartley, and D.~Campbell, ``Deep declarative networks: A new
  hope,'' \emph{arXiv preprint arXiv:1909.04866}, 2019.

\bibitem{boyd2004convex}
S.~Boyd, S.~P. Boyd, and L.~Vandenberghe, \emph{Convex optimization}.\hskip 1em
  plus 0.5em minus 0.4em\relax Cambridge university press, 2004.

\bibitem{diff-layers}
A.~Agrawal, B.~Amos, S.~Barratt, S.~Boyd, S.~Diamond, and J.~Z. Kolter,
  ``Differentiable convex optimization layers,''
  \url{https://locuslab.github.io/2019-10-28-cvxpylayers/}.

\bibitem{goodfellow2016deep}
I.~Goodfellow, Y.~Bengio, and A.~Courville, \emph{Deep learning}.\hskip 1em
  plus 0.5em minus 0.4em\relax MIT press, 2016.

\bibitem{4801507}
L.~P. Qian, Y.~J. Zhang, and J.~Huang, ``{MAPEL:} {A}chieving global optimality
  for a non-convex wireless power control problem,'' \emph{IEEE Trans. on
  Wireless Commun.}, vol.~8, no.~3, pp. 1553--1563, 2009.

\bibitem{shi2014group}
Y.~Shi, J.~Zhang, and K.~B. Letaief, ``Group sparse beamforming for green
  cloud-ran,'' \emph{IEEE Trans. on Wireless Commun.}, vol.~13, no.~5, pp.
  2809--2823, 2014.

\bibitem{ioffe2015batch}
S.~Ioffe and C.~Szegedy, ``Batch normalization: Accelerating deep network
  training by reducing internal covariate shift,'' in \emph{Intl. Conf. on
  machine learning}.\hskip 1em plus 0.5em minus 0.4em\relax PMLR, 2015, pp.
  448--456.

\bibitem{hinton2012improving}
G.~E. Hinton, N.~Srivastava, A.~Krizhevsky, I.~Sutskever, and R.~R.
  Salakhutdinov, ``Improving neural networks by preventing co-adaptation of
  feature detectors,'' \emph{arXiv preprint arXiv:1207.0580}, 2012.

\bibitem{paszke2019pytorch}
A.~Paszke, S.~Gross, F.~Massa, A.~Lerer, J.~Bradbury, G.~Chanan, T.~Killeen,
  Z.~Lin, N.~Gimelshein, L.~Antiga \emph{et~al.}, ``Pytorch: An imperative
  style, high-performance deep learning library,'' \emph{Advances in neural
  information processing systems}, vol.~32, 2019.

\bibitem{ecos}
A.~Domahidi, E.~Chu, and S.~Boyd, ``{ECOS}: {A}n {SOCP} solver for embedded
  systems,'' in \emph{European Control Conf. (ECC)}, 2013, pp. 3071--3076.

\bibitem{diamond2016cvxpy}
S.~Diamond and S.~Boyd, ``{CVXPY}: {A} {P}ython-embedded modeling language for
  convex optimization,'' \emph{Journal of Machine Learning Research}, vol.~17,
  no.~83, pp. 1--5, 2016.

\bibitem{mosek}
\BIBentryALTinterwordspacing
M.~ApS, \emph{MOSEK Optimizer API for Python 9.3.20}, 2019. [Online].
  Available: \url{https://docs.mosek.com/9.3/pythonapi/index.html}
\BIBentrySTDinterwordspacing

\bibitem{lee2021robust}
W.~Lee and K.~Lee, ``Robust transmit power control with imperfect csi using a
  deep neural network,'' \emph{IEEE Trans. on Vehicular Technology}, vol.~70,
  no.~11, pp. 12\,266--12\,271, 2021.

\bibitem{ben2009robust}
A.~Ben-Tal, L.~El~Ghaoui, and A.~Nemirovski, \emph{Robust optimization}.\hskip
  1em plus 0.5em minus 0.4em\relax Princeton university press, 2009, vol.~28.

\bibitem{liu2022deep}
Z.~Liu, Y.~Yang, F.~Gao, T.~Zhou, and H.~Ma, ``Deep unsupervised learning for
  joint antenna selection and hybrid beamforming,'' \emph{IEEE Transactions on
  Communications}, vol.~70, no.~3, pp. 1697--1710, 2022.

\bibitem{kim2021deep}
M.~Kim, H.~Lee, H.~Lee, and I.~Lee, ``Deep learning based resource assignment
  for wireless networks,'' \emph{IEEE Communications Letters}, vol.~25, no.~12,
  pp. 3888--3892, 2021.

\end{thebibliography}
\begin{IEEEbiography}
	[{\includegraphics[width=1in,height=1.35in]{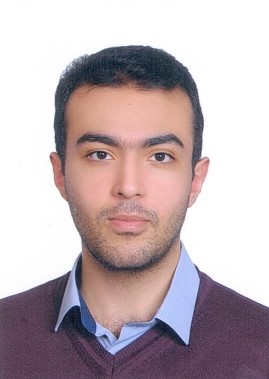}}]
	{Mehrazin Alizadeh} obtained his M.Sc. degree in Computer Engineering from York University, Toronto, ON, Canada, in 2022, after receiving a B.Sc. degree in Electrical Engineering from the University of Tehran, Tehran, Iran, in 2020. His research is fundamentally centered around the applications of deep learning in addressing resource allocation problems in wireless communications networks. In particular, his work involves leveraging deep neural networks to design novel solvers for optimization problems within wireless networks, offering significant computational advantages over traditional methodologies.
\end{IEEEbiography}

\begin{IEEEbiography}
	[{\includegraphics[width=1.1in,height=1.35in]{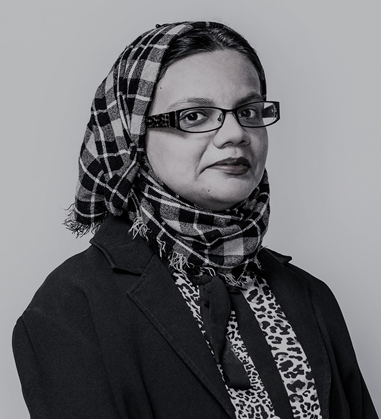}}]
	{Hina Tabassum} is currently a faculty member in the  Lassonde School of Engineering, York University, Canada. Prior to that, she was a postdoctoral research associate at University of Manitoba, Canada. She received her PhD degree from King Abdullah University of Science and Technology (KAUST) in 2013. 
	She received Lassonde Innovation Award in 2023, N2Women: Rising Stars in Computer Networking and Communications in 2022, and listed in the Stanford's list of the World’s Top Two-Percent Researchers in 2021 and 2022. She is the founding chair of a special interest group on THz communications in IEEE Communications Society (ComSoc) - Radio Communications Committee (RCC). She is a Senior member of IEEE and registered Professional Engineer in the province of Ontario, Canada. Currently, she is serving as an Area Editor in IEEE Open Journal of Communications Society and Associate Editor in IEEE IoT Magazine, IEEE Transactions on Communications, IEEE Communications Letters, IEEE Transactions on Green Communications, IEEE Communications Surveys and Tutorials.  Her research interests include stochastic modeling and optimization of wireless networks including vehicular, aerial, and satellite networks, millimeter and terahertz communication networks.
\end{IEEEbiography}
\end{document}